\documentclass[a4paper,reqno]{amsart}

\usepackage{graphicx}

\usepackage{subfigure}
\usepackage{float}
\usepackage{xcolor}
\usepackage{cite}

\usepackage{amssymb}
\usepackage{amsthm}
\usepackage{mathtools}
\usepackage{mathrsfs}
\newcommand*{\sM}{{\mathscr{M}}}
\newcommand*{\sS}{{\mathcal{S}}}
\newcommand*{\sT}{{\mathcal{T}}}
\newcommand*{\sV}{{\mathcal{V}}}
\newcommand*{\DiracD}{\mathscr{D}}
\newcommand*{\ii}{\mathrm{i}}
\newcommand*{\CC}{\mathbb{C}}
\newcommand*{\VV}{\mathbb{V}}
\newcommand*{\dd}{\mathrm{d}}
\newcommand*{\F}{\mathcal{F}}
\newcommand*{\dotF}{\mathcal{F}_t}

\usepackage{enumerate}

\usepackage[ colorlinks = true,
             linkcolor = blue,
             urlcolor  = blue,
             citecolor = red,
             anchorcolor = green,
             hyperindex = true,
           ]{hyperref}

\usepackage{cleveref}

\graphicspath{
{figures/}
{data/}
}

\theoremstyle{plain}

\theoremstyle{definition}

\let\oldsqrt\sqrt
\def\sqrt{\mathpalette\DHLhksqrt}
\def\DHLhksqrt#1#2{%
\setbox0=\hbox{$#1\oldsqrt{#2\,}$}\dimen0=\ht0
\advance\dimen0-0.2\ht0
\setbox2=\hbox{\vrule height\ht0 depth -\dimen0}%
{\box0\lower0.4pt\box2}}

\title[A spectral method for half-integer spin fields]{A spectral method for half-integer spin fields based on spin-weighted spherical harmonics}

\author[F. Beyer]{Florian Beyer}
\address{Department of Mathematics and Statistics, University of Otago, PO Box 56, Dunedin 9010, New Zealand}
\email{fbeyer@maths.otago.ac.nz}
\author[B. Daszuta]{Boris Daszuta}
\address{Department of Mathematics and Statistics, University of Otago, PO Box 56, Dunedin 9010, New Zealand}
\email{bdaszuta@maths.otago.ac.nz}
\author[J. Frauendiener]{J\"org Frauendiener}
\address{Department of Mathematics and Statistics, University of Otago, PO Box 56, Dunedin 9010, New Zealand}
\email{joergf@maths.otago.ac.nz}
\numberwithin{equation}{section}

\begin{document}
\newlength\figureheight
\newlength\figurewidth

\begin{abstract}
  We present a new spectral scheme for analysing functions of
  half-integer spin-weight on the $2$-sphere and demonstrate the
  stability and convergence properties of our implementation.  The
  dynamical evolution of the Dirac equation on a manifold with spatial
  topology of $\mathbb{S}^2$ with a pseudo-spectral method is also
  demonstrated.
\end{abstract}
\maketitle

\section{Introduction}

In a recent paper~\cite{numericalEvolutions2014Beyer} we presented a spectral
method for tensorial partial differential equations on geometries with a
spherical component. We showed how to implement the `eth'-formalism based on
spin-weighted spherical harmonics following the work by Newman, Penrose and
others~\cite{Goldberg:1967vm,newman_note_1966}. The restriction to tensorial
equations implied the use of spin-weighted functions with integer
spin-weights. In the present work we extend our method to also include functions
with half-integer spin-weight on spherical geometries.  In particular, we show
how the evolution of dynamical half-integer spin fields can be accomplished. Our
motivation for doing so stems mainly from General Relativity: spinorial fields
describing elementary particles such as electrons or neutrinos are fundamental
sources for the Einstein equations and are studied for instance as test fields
on Kerr-Newman backgrounds, see~\cite{Finster:2006} and references
therein. However, there are also interesting developments in condensed matter
physics in relation to the description of
graphene~\cite{Neto2009TheElectronicProperties,Vozmediano2010GaugeFields} for
which our method might be relevant. The numerical treatment of the Dirac field
is particularly interesting as its evolution can be rather peculiar and
sometimes even counter-intuitive as visualised by
Thaller~\cite{Thaller:2004uh,Thaller:2005vi}. 

It is well known that when working with $\mathbb{S}^2$ (and other compact
geometries) coordinate singularities can lead to instabilities that spoil the
accuracy of numerical schemes. These issues can be avoided by working with
spectral methods where coordinate-independent manipulation of expressions in
terms of a well-defined basis of functions upon which the action of differential
operators reduces to algebraic manipulation is possible
\cite{boyd2001chebyshev,grandclement2009spectral}. Thus, issues of coordinate
singularities are dealt with automatically by the method.

Our spectral method makes use of the spin-weighted spherical harmonics (SWSH)
\cite{newman_note_1966,Goldberg:1967vm,Penrose:1984tf,Penrose:1986tf} which can
be shown to be equivalent to the well-known Wigner $D$-functions
\cite{AUnified1986Dray,Sugiura:1990vj,Sakurai:1994modern} and thus provide a
complete, orthonormal basis for $L^2(SU(2))$. In the description of half-integer
spin fields, we have found it convenient to work directly with the half-integer
SWSH due to the particularly simple action of the associated $\eth$ and $\eth'$
differential operators which may be thought of as constituting covariant
derivative operators on the sphere \cite{Goldberg:1967vm}.  As the action of the
aforementioned operators on the SWSH reduces to simple algebraic manipulation we
may translate PDE systems into coupled (infinite dimensional) ODE systems and
thus construct a spectral evolution scheme for an initial value problem (IVP) of
interest.

Our work extends that of \cite{Huffenberger:2010hh,numericalEvolutions2014Beyer}
into a spectral algorithm for spin-weighted spherical harmonics (SWSH) with
half-integer spin-weight $s$. The method we present inherits several desirable
properties from~\cite{Huffenberger:2010hh,numericalEvolutions2014Beyer}. In
particular it is theoretically exact if a minimum number of grid points are used
at a given band-limit $L$. Furthermore the same algorithmic complexity of
$\mathcal{O}(L^3)$ is achievable for spectral transformations. In brief,
\cite{Huffenberger:2010hh,numericalEvolutions2014Beyer} seek to calculate values
for the integer SWSH over $\mathbb{S}^2$ by mapping the sphere into the
$2$-torus (thus allowing for Fast Fourier Transforms to be used) and relate the
SWSH to the reduced Wigner $d$-functions evaluated at $\pi/2$. For a short
summary of past work related to the integer case see
\cite{numericalEvolutions2014Beyer}.  The approach to the half-integer case is
in principle similar, however suitable modification of the $2$-torus map must be
made in order to account for the periodicity of spinor fields. In addition, the
recursion relation \cite{Trapani:2006he} that allows for evaluation of the
Wigner $d$-functions must be modified.

As a toy model we numerically explore the dynamics associated with a Dirac
equation on a 2-dimensional manifold with spatial topology of $\mathbb{S}^2$ as
an IVP. This will allow for a test of the spin-weighted spectral transformations
we present in the context of evolution equations. Common techniques for treating
the numerical problem of the Dirac equation consist of: FD schemes formulated on
a flat-lattice in configuration space
\cite{StaggeredGrid2014Hammer,SingleCone2014Hammer}, on a grid within a
finite-volume in momentum-space \cite{NumericalTreat1996Momberger} and using
methods based on the split-step operator technique
\cite{QuantumDynamics2004Mocken,FFTsplit2008Mocken,NumericalSoln2012Gourdeau,ASplitStep2014Gordeau}.
Particular to the FD approach special care must be taken so as to avoid the
Fermion-doubling problem \cite{ANoGo1981Nielsen}. Elimination of spurious modes
introduced to the solution may be accomplished by means of nonlocal
approximation for the spatial derivative operator
\cite{Eliminating1982Stacey,FiniteDifference2008Tworzydlo} or by staggered-grid
schemes \cite{StaggeredGrid2014Hammer,SingleCone2014Hammer}. As remarked upon in
\cite{SingleCone2014Hammer} the issue of spurious modes is not particular to the
Dirac equation but can occur whenever a symmetric FD approximant is used for a
first derivative on a uniform grid.  We seek to avoid the above issue entirely
by making use of the global approximation to functions and derivative operators
that spectral methods provide \cite{hesthaven2007spectral}.

The equations of motion (EOM) that we derive for the Dirac equation in the
aforementioned geometric setting result in a coupled system of
variable-coefficient linear hyperbolic PDEs. When performing a SWSH
decomposition this results in product terms that must be further simplified.
From past experience \cite{numericalEvolutions2014Beyer} we have found that
while it is possible to make use of Clebsch-Gordan expansions (see
sec.~\ref{sec:swshsum}), i.e., working entirely in the space of coefficients, it
is far more convenient to instead work with a pseudo-spectral method (see
sec.~\ref{sec:psmethod}), and thus we follow the latter approach in this work.

This paper is structured as follows: In sec.~\ref{sec:swshsum} we recall basic
properties of the half-integer SWSH together with the $\eth$-formalism. In
sec.~\ref{sec:fwdxform} and \ref{sec:bwdxform} respectively we discuss the
forward and backward spectral transformations of spin-weighted functions on the
sphere.  In sec.~\ref{sec:wigrec} we present the required modification to the
recursion relation for computing the reduced Wigner functions at $\pi/2$.
Subsequently, in sec.~\ref{sec:consischk} we perform consistency checks on our
implementation of the algorithm. Error pairs and the property of exponential
convergence are analysed.  In sec.~\ref{sec:Dirac} we proceed by numerically
solving the $2+1$ dimensional Dirac equation as an IVP on a curved geometry,
specifying to cases with spatial topology of $\mathbb{S}^2$.  We find EOM
adapted to the spin-weighted formalism.  In sec.~\ref{sec:psmethod} we briefly
recall the pseudo-spectral method. In sec.~\ref{sec:numericalEOM} we construct
numerical solutions to the EOM and inspect convergence properties of the
numerical solutions obtained together with conserved
currents. Sec.~\ref{sec:conc} concludes.

\section{SWSH summary}\label{sec:swshsum}

In this section we briefly summarise key properties of the spin-weighted spherical harmonics (SWSH) that we will make use of in the presentation of our half-integer spectral algorithm and in the pseudo-spectral method for solution of dynamical systems. For further details we refer the reader to \cite{Goldberg:1967vm,newman_note_1966,Penrose:1984tf,Penrose:1986tf,AUnified1986Dray,numericalEvolutions2014Beyer}.

Geometrically, the spin-weighted functions are sections of certain line bundles over the 2-sphere $\mathbb{S}^2$. They have a representation in terms of ordinary functions over patches of $\mathbb{S}^2$ which depend on the choice of coordinates and the choice of an orthonormal frame at points of the patch. The behaviour of this representation under changes of the frame is captured by the spin-weight assigned to the function. Here, we will deal only with representations of the spin-weighted functions given in terms of the standard polar coordinates $(\vartheta,\varphi)$ and the orthonormal frame defined in terms of the coordinate derivative vectors
\begin{equation}
  \label{eq:cordframe}
  {e}_\vartheta = \partial_\vartheta, \quad 
  {e}_\varphi = \frac1{\sin\vartheta} \partial_\varphi.
\end{equation}
Equivalently, these vectors can be obtained as real and imaginary parts of the complex linear combination
\[
{m} = \frac{1}{\sqrt2} \left( {e}_\vartheta - \ii {e}_\varphi \right) = 
\frac{1}{\sqrt2} \left( \partial_\vartheta - \frac{\ii}{\sin\vartheta} \partial_\varphi \right) 
\]
which, together with its complex conjugate, satisfies the orthonormality relations
\[
{m}\cdot{m} =0, \quad
{\overline m}\cdot{\overline m}=0, \quad
{m}\cdot{\overline m}=1
\]
at all points of $\mathbb{S}^2$ covered by the polar coordinates with respect to the standard metric on $\mathbb{S}^2$.

Every smooth function ${}_{s}f$ on $\mathbb{S}^2$ with spin-weight $s$ can be expressed as:
\begin{equation}\label{eq:funcs2sphTrunc}
{}_{s} f(\vartheta,\varphi)
=
\lim_{L\rightarrow \infty}\sum^L_{l=|s|}\sum^{l}_{m=-l} {{}_sa_{lm}} \,{{}_sY_{lm}(\vartheta,\varphi)},
\end{equation}
where the ${}_sY_{lm}$ are the SWSH, which form a complete orthonormal basis for the complex vector space of spin-weight $s$ functions.

Recall that the action of the $\eth$ and $\eth'$ differential operators on ${}_sf$ serves to raise and lower the spin-weight respectively; when 
viewed as maps with respect to polar coordinates we have:
\begin{align}\label{eq:actionEth}
\eth:{}_sf&\rightarrow {}_{s+1}\tilde{f}, &&\eth\left[{}_sf\right]=
\partial_\vartheta[{}_sf]-\ii\csc\vartheta\,\partial_\varphi[{}_sf] - s\cot\vartheta {}_sf,\\\label{eq:actionEthp}
\eth':{}_sf&\rightarrow {}_{s-1}\tilde{f}, &&\eth'\left[{}_sf\right]=
\partial_\vartheta[{}_sf]+\ii\csc\vartheta\,\partial_\varphi[{}_sf] + s\cot\vartheta {}_sf.
\end{align}
The SWSH may be written explicitly as:
\begin{equation}\label{eq:swshDefn}
{}_{s}Y_{lm}(\vartheta,\varphi) = \sqrt{\frac{2l+1}{4\pi}}\exp(\ii m\varphi)\,d^l_{sm}(\vartheta),
\end{equation}
where $d^l_{mn}(\vartheta)$ is the reduced Wigner $d$-function:
\begin{equation}
\label{eq:dwigreppol}
\begin{multlined}
d^l_{mn}(\vartheta)
	= \sum^{\mathrm{min}(l+m,l-n)}_{r=\mathrm{max}(0,m-n)} (-1)^{r-m+n}
		\frac{\sqrt{(l+m)!(l-m)!(l+n)!(l-n)!}}{r!(l+m-r)!(l-r-n)!(r-m+n)!} \times\\
\times		\cos^{2l-2r+m-n}\left(\frac{\vartheta}{2} \right) \sin^{2r-m+n}\left(\frac{\vartheta}{2}\right).
              \end{multlined}
\end{equation}
Under complex conjugation the SWSH satisfy:
\begin{equation}\label{eq:swshconj}
\overline{{}_sY_{lm}}=(-1)^{s-m}\,{}_{-s}Y_{l,-m} \Longleftrightarrow {}_sY_{lm}=(-1)^{s-m}\,\overline{{}_{-s}Y_{l,-m}}.
\end{equation}
The SWSH satisfy the orthonormality relation:
\begin{equation}\label{eq:sYlmOrth}
\left\langle {{}_sY_{l_1m_1}},\,{{}_sY_{l_2m_2}}\right\rangle
=
\int_0^{2\pi}\int_0^\pi
{}_sY_{l_1m_1}(\vartheta,\varphi) \overline{{}_sY_{l_2m_2}(\vartheta,\varphi)} 
\sin\vartheta\,d\vartheta\,d\varphi
=
\delta_{l_1l_2}\delta_{m_1m_2}.
\end{equation}
A particularly useful property the SWSH possess is that under the
action of $\eth$ and $\eth'$ the action of Eq.~\eqref{eq:actionEth} and
Eq.~\eqref{eq:actionEthp} reduces to a "ladder algebra":
\begin{align}\label{eq:ladder1}
\eth\left[{}_sY_{lm}(\vartheta,\varphi) \right] &=
 -\sqrt{(l-s)(l+s+1)}{}_{s+1}Y_{lm}(\vartheta,\varphi),\\\label{eq:ladder2}
\eth'\left[{}_sY_{lm}(\vartheta,\varphi) \right] &=
 \sqrt{(l+s)(l-s+1)}{}_{s-1}Y_{lm}(\vartheta,\varphi),
\end{align}
which we will exploit when performing decompositions of dynamical equations, allowing for a map from a PDE system to an 
(infinite dimensional) ODE system (see sec.~\ref{sec:Dirac}). In addition to this we have the commutator expression:
\begin{equation}
\left[\eth,\eth'\right]\,{}_sY_{lm}(\vartheta,\varphi)=-2s\,{}_sY_{lm}(\vartheta,\varphi).
\end{equation}

A product of two SWSH with spin-weights $s_1$ and $s_2$ is a function of spin-weight $s_1+s_2$ and, therefore, it can be written as a finite linear combination of SWSH 
\begin{equation}\label{eq:prodSoln}
	{}_{s_{1}}Y_{l_{1},m_{1}}(\vartheta,\,\varphi){}_{s_{2}}Y_{l_{2},m_{2}}(\vartheta,\,\varphi)
	=
	\sum_{l\in\Lambda'}\mathcal{C}_l(s_1,l_1,m_1;\,s_2,l_2,m_2)
	{}_{(s_{1}+s_{2})}Y_{l,(m_{1}+m_{2})}(\vartheta,\,\varphi),
\end{equation}
where $\Lambda':=\{\max(|l_1-l_2|,\,|s_1+s_2|,\,|m_1+m_2|),\,\dots,\,l_1+l_2\}$ and the coefficients are related to the usual Clebsch-Gordan coefficients, see e.g.~\cite{Sakurai:1994modern,numericalEvolutions2014Beyer}.

\subsection{Forward transformation}\label{sec:fwdxform}
We now describe our numerical algorithm for evaluation of the forward transform $\mathcal{F}: {}_sf\mapsto ({}_sa_{lm})$. 
As a first step we introduce the notation $\Delta_{m n}^l:=d_{m n}^l\left(\pi/2\right)$ which allows for the rewriting of Eq.~\eqref{eq:dwigreppol} as~\cite{Risbo:1996iy}:
\begin{equation}
\label{eq:wignerdDecDelta} 
d_{m n}^l(\vartheta)=\ii^{m-n}\sum_{q=-l}^{l}\Delta_{q m}^le^{-\ii q\vartheta}\Delta_{q n}^l,
\end{equation} 
following from a factoring of rotations \cite{Trapani:2006he}. In particular, note that $d^l_{mn}(\vartheta)=\overline{d^l_{mn}(\vartheta)}$ 
(cf. Eq.~\eqref{eq:dwigreppol}). 
We defer the details of how the $\Delta$ elements are calculated together with their symmetry properties to sec.~\ref{sec:wigrec}.

Define the functional:
\begin{equation}\label{eq:ImnQuadPr}
I_{mn}\left[{}_sf(\vartheta,\varphi) \right]:=\int^{2\pi}_0\int^\pi_0 e^{-\ii m\vartheta}e^{-\ii n\varphi}{}_s f(\vartheta,\varphi) \sin\vartheta \,d\vartheta d\varphi.
\end{equation}
This integral can be evaluated exactly -- we now describe our method which is based on \cite{Huffenberger:2010hh}. 
Combining \cref{eq:ImnQuadPr,eq:wignerdDecDelta,eq:swshDefn,eq:funcs2sphTrunc} results in:
\begin{align}
{}_sa_{lm} &= \ii^{s-m} \sqrt{\frac{2l+1}{4\pi}}\sum_{q=-l}^l \Delta^l_{qs} I_{qm} \Delta^l_{qm},\\
\label{eq:salmfwd}
&= \ii^{s-m} \sqrt{\frac{2l+1}{4\pi}}\sum_{q=1/2}^l \Delta^l_{qs} J_{qm} \Delta^l_{qm},
\end{align}
where
\begin{equation}
J_{qm}:= I_{qm}+(-1)^{2l+s+m}I_{-q,m}.
\end{equation}

\begin{figure}[ht]
	\centering
	\includegraphics[height=0.25\textheight]{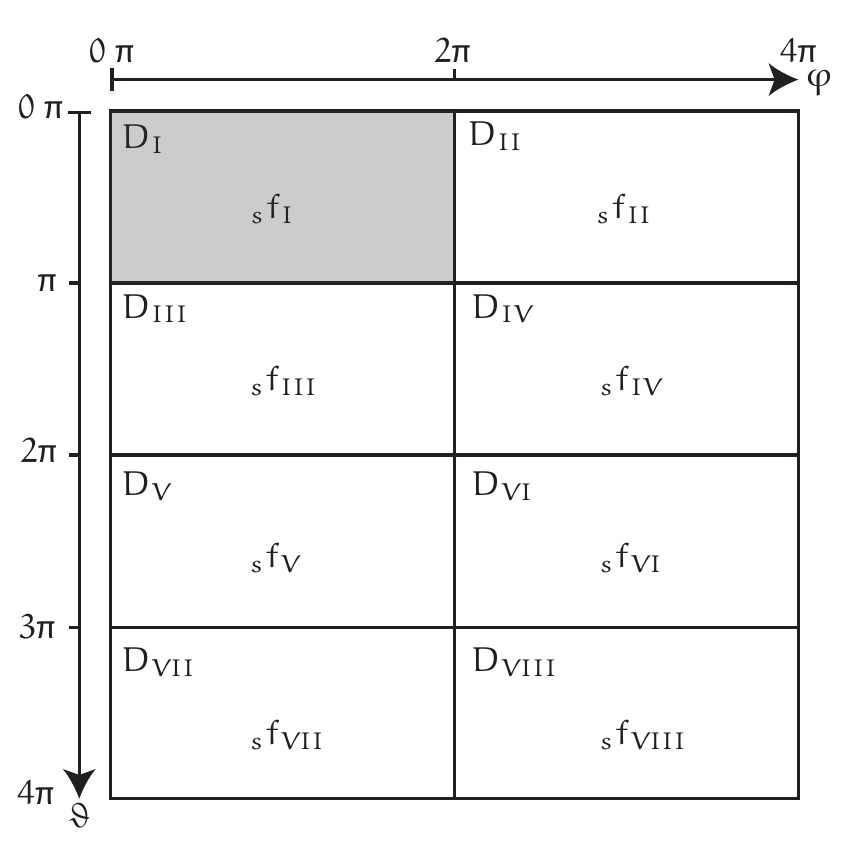}
	\caption{
	Domains for function extension. The shaded region corresponds to data that are to be extended to unshaded regions.
	}
	\label{fig:extDom}
\end{figure}

In order to make use of existing FFT (fast Fourier transform) algorithms in the evaluation of the quadrature in Eq.~\eqref{eq:ImnQuadPr}, we must extend a spin-weighted function ${}_sf$ with data sampled on the domain $D:=\{(\vartheta,\varphi)|\vartheta\in[0,\pi],\varphi\in[0,2\pi]\}$ to a $4\pi$ periodic function on $\tilde{D}:=\{(\vartheta,\varphi)|\vartheta\in[0,4\pi),\varphi\in[0,4\pi)\}$. This is accomplished using the intrinsic symmetries of the ${}_sY_{lm}(\vartheta,\varphi)$ together with the expansion ansatz of Eq.~\eqref{eq:funcs2sphTrunc}. 
For convenience we define the domains $D_\mathrm{I},\,\dots,D_\mathrm{VIII}$ (see Fig.~\ref{fig:extDom}). 
We have $D=D_\mathrm{I}$ upon which ${}_sf(\vartheta,\varphi)={}_sf_\mathrm{I}(\vartheta,\varphi)$. For $D_\mathrm{II}$ put ${}_sf_\mathrm{II}(\vartheta,\varphi)=-{}_sf_\mathrm{I}(\vartheta,\varphi-2\pi)$. Next, define ${}_sg:D_\mathrm{I}\cup D_\mathrm{II}\rightarrow \mathbb{C}$ by
\begin{equation}
{}_s g(\vartheta,\varphi):=
	\begin{cases}
		{}_sf_\mathrm{I}(\vartheta,\varphi) & (\vartheta,\varphi)\in D_\mathrm{I}\\
		{}_sf_\mathrm{II}(\vartheta,\varphi) & (\vartheta,\varphi) \in D_\mathrm{II}
	\end{cases}.
\end{equation}
Now, define ${}_s h:D_\mathrm{I}\cup D_\mathrm{II} \cup D_\mathrm{III} \cup D_\mathrm{IV} \rightarrow \mathbb{C}$ by
\begin{equation}
{}_s h(\vartheta, \varphi):=
	\begin{cases}
		{}_s g(\vartheta,\varphi) & (\vartheta,\varphi)\in D_\mathrm{I} \cup D_\mathrm{II}\\
		(-1)^{s+1} {}_s g(2\pi-\vartheta,(\varphi-\pi)\mathrm{mod}4\pi ) & (\vartheta,\varphi)\in D_\mathrm{III} \cup D_\mathrm{IV}
	\end{cases}
\end{equation}
Finally, we arrive at the extended function ${}_sF:\tilde{D}\rightarrow\mathbb{C}$ defined by
\begin{equation}
{}_s F(\vartheta,\varphi):=
	\begin{cases}
		{}_sh(\vartheta,\varphi) & (\vartheta,\varphi)\in D_\mathrm{I}\cup D_\mathrm{II}\cup D_\mathrm{III} \cup D_\mathrm{IV}\\
		-{}_sh(\vartheta-2\pi,\varphi) & (\vartheta,\varphi) \in D_\mathrm{V}\cup D_\mathrm{VI} \cup D_\mathrm{VII}\cup D_\mathrm{VIII}
	\end{cases}
\end{equation}
The smooth function ${}_sF(\vartheta,\varphi)$ is $4\pi$ periodic in $\vartheta$ and $\varphi$, thus we may write:
\begin{equation}\label{eq:expaExt}
{}_sF(\vartheta,\varphi)=\sum_{k=0}^{K}\sum_{m=0}^M {}_sF_{km} e^{\ii
  k\frac{2\pi}{4\pi}\vartheta} e^{\ii m\frac{2\pi}{4\pi}\varphi}.
\end{equation}
In order to evaluate~\eqref{eq:ImnQuadPr} we now identify
\begin{align}
I_{mn}\left[{}_sf(\vartheta,\varphi) \right] \equiv I_{mn}\left[{}_sF(\vartheta,\varphi)\right] &=
\int_0^{2\pi} \int_0^\pi e^{-\ii m\vartheta/2}e^{-\ii n\varphi/2}{}_sF(\vartheta,\varphi)\sin\vartheta\,d\vartheta d\varphi,
\end{align}
where we can extend the indices to all integers, i.e., $m,n \in \mathbb{Z}$. This is permissible as both functions are equal in the region of integration. Upon substitution with Eq.~\eqref{eq:expaExt} we find:
\begin{align}\label{eq:quadspec}
I_{mn}\left[{}_sF(\vartheta,\varphi)\right] &=
\sum_{i,j=0} {}_s F_{ij} \;W_\vartheta(i-m)W_\varphi(j-n),
\end{align}
with
\begin{align}
W_\vartheta(\tau):&=\int_0^\pi e^{\ii\tau\vartheta/2}\sin\vartheta\,d\vartheta =
	\begin{cases}
		\pm \ii\frac{\pi}{2} &\tau=\pm 2\\
		4\frac{1+\ii^\tau}{4-\tau^2} & \tau\in\mathbb{Z}\setminus\{\pm2\}
	\end{cases},\\
W_\varphi(\rho):&=\int_0^{2\pi} e^{\ii\rho\varphi/2}\,d\varphi =
	\begin{cases}
		2\pi & \rho=0\\
		\frac{2\ii}{\rho}(1-(-1)^\rho) & \rho\in\mathbb{Z}\setminus\{0\}
	\end{cases}.
\end{align}
Equation \eqref{eq:quadspec} amounts to a two-fold discrete convolution in spectral space. By the convolution theorem, this implies that we may equivalently consider pointwise multiplication of the inverse transforms of $W_\vartheta$ and $W_\varphi$ with ${}_sF$. 
Let $N_\vartheta$ and $N_\varphi$ denote the number of grid points over $\vartheta$ and $\varphi$ on the containerization of $\mathbb{S}^2$ that 
the function ${}_sf(\vartheta,\varphi)$ is to be sampled at. Upon extension to the doubly $4\pi$ periodic domain (see Fig.~\ref{fig:extDom}) we 
work instead with the extended function ${}_sF(\vartheta,\varphi)$. Furthermore, in order to simplify 
numerical construction of the extension, the number of samples on the extended domain is taken to be $N_\vartheta'=4N_\vartheta-3$ and $N_\varphi'=2N_\varphi-1$. Correspondingly, the spatial sampling intervals are now given by $\Delta\vartheta=\frac{4\pi}{N_\vartheta'-1}$ and $\Delta\varphi=\frac{4\pi}{N_\varphi'-1}$. In order to satisfy the Nyquist condition, such that spurious aliasing does not occur, we impose 
$N_\vartheta=N_\varphi=2(L+2-1/2)+1$, where $L$ is the harmonic that the function ${}_sf(\vartheta,\varphi)$ is band-limited to\footnote{ 
More efficient samplings on the sphere may be possible \cite{McEwen:2011:NovelSampling} however the overall asymptotic complexity of our proposed algorithm will not be 
affected and thus we do not explore this issue here.}.

Under these choices a convenient form of the quantities discussed above that is directly amenable to numerical work is given by:
\begin{align}
\tilde{W}_\vartheta(\gamma) &:= 
\sum_{\delta=0}^{N_\vartheta'-2} e^{\ii\gamma\delta\Delta\vartheta/2} W_\vartheta\left(\frac{N_\vartheta'-1}{2}-\delta\right) &
\gamma=& 0,\dots N_\vartheta'-2\\
\tilde{W}_\varphi(\varepsilon) &:= 
\sum_{\zeta=0}^{N_\varphi'-2} e^{\ii\varepsilon \zeta \Delta\varphi/2} W_\varphi\left(\frac{N_\varphi'-1}{2}-\zeta\right) &
\varepsilon=& 0,\dots N_\varphi'-2
\end{align}
together with:
\begin{align}\nonumber
I_{\mu\nu}\left[{}_sF(\vartheta,\varphi)\right] =&
\frac{1}{(N_\vartheta'-1)(N_\varphi'-1)}
		\sum_{\gamma=0}^{N_\vartheta'-2}\sum_{\varepsilon=0}^{N_\varphi'-2}
			e^{-\ii\mu\gamma \Delta \vartheta/2}
			e^{-\ii\nu\varepsilon\Delta\varphi/2}
\\\label{eq:Imunudft}
&		\times
				\tilde{W}_\vartheta(\gamma)
				\tilde{W}_\varphi(\varepsilon)
				{}_sF(\gamma,\varepsilon).
\end{align}
where $\mu\in{0,\dots,\,N_\vartheta'-2}$, $\nu\in\{0,\dots,\,N_\varphi'-2\}$.
Note that in this approach, we must discard all values of $I_{\mu\nu}$ for which
$\mu$ and $\nu$ are even integers.  In order to recover $I_{mn}$ from
Eq.\eqref{eq:Imunudft} we take $\mu=\frac{1}{2}(N_\vartheta'-4m-1)$ and
$\nu=\frac{1}{2}(N_\varphi'-4n-1)$.  Overall we find an algorithmic complexity
of $\mathcal{O}(L^3)$ as the integrals $I_{\mu\nu}$ may be evaluated exactly in
$\mathcal{O}(L^2\log L)$ operations by performing a two-dimensional FFT and each
component of $\Delta^l_{mn}$ (as required by Eq.~\eqref{eq:salmfwd}) can be
computed using $\mathcal{O}(1)$ floating point evaluations -- see
sec.~\ref{sec:wigrec} and also \cite{Trapani:2006he,Huffenberger:2010hh}.  We
remark that if the analysis of strictly real data is desired then it is possible
to attain a linear increase in the execution speed of transformations
(specifically the FFT component by a factor of approximately $2$).  However, as
we are primarily interested in applying transformations for the solution of
equations of motions that govern the dynamics of complex fields we do not
explore this further.

\subsection{Backward transformation}\label{sec:bwdxform}
We now describe the algorithm for evaluation of the backward (inverse) transform $\mathcal{F}{}^{-1}:({}_sa_{lm})\mapsto{}_sf$.  
The backward spherical harmonic transform maps the expansion coefficients ${}_sa_{lm}$, for $|s|\leq l \leq L$, 
to a function on (a dense subset of)
$\mathbb{S}^2$.  Because we are working with band-limited functions we can, at the analytical level, 
perfectly reconstruct the original function. To this end, Eq.~\eqref{eq:funcs2sphTrunc} must be evaluated. 
As the inverse transform does not contain integrals, issues of quadrature accuracy do not arise.
Define:
\begin{equation}
K_{mn}\left[{}_sa_{ln}\right]:= \ii^{s-n} \sum_{l=|s|}^L \sqrt{\frac{2l+1}{4\pi}} \Delta^l_{-m,s}{}_sa_{ln}\Delta^l_{-m,n}
\end{equation}
which allows for:
\begin{equation}\label{eq:reconstructFunc}
{}_sf(\vartheta,\varphi) =
\sum_{m=-\tfrac12(N_\vartheta'-1)+\tfrac12}^{\tfrac12(N_\vartheta'-1)-\tfrac12}
\sum_{n=-\tfrac12(N_\varphi'-1)+\tfrac12}^{\tfrac12(N_\varphi'-1)-\frac12}
e^{\ii m\vartheta} e^{\ii n\varphi} K_{mn}.
\end{equation}
Use of Eq.~\eqref{eq:delsym1} permits a rewriting of Eq.~\eqref{eq:reconstructFunc} as:
\begin{equation}
K_{mn}\left[{}_sa_{ln}\right]= 
	\begin{cases}
		\ii^{s-n} \sum_{l=|s|}^L \sqrt{\frac{2l+1}{4\pi}} \Delta^l_{-m,s}{}_sa_{ln}\Delta^l_{-m,n} & m\geq \frac{1}{2}\\
		\ii^{s-n} \sum_{l=|s|}^L \sqrt{\frac{2l+1}{4\pi}} (-1)^{2l+s+n}\Delta^l_{m,s}{}_sa_{ln}\Delta^l_{m,n} & m\leq -\frac{1}{2}
	\end{cases},
\end{equation}
which allows for a reduction in computation time.
In Eq.~\eqref{eq:reconstructFunc} we may use an FFT directly, discarding values of ${}_sf(\vartheta,\varphi)$ for which $\vartheta>\pi$ and $\varphi>2\pi$.
If the input data to the FFT library is Hermitean then another linear increase ($\sim2$) in the execution speed of a transform is possible 
however, this again does not change the overall algorithmic complexity $\mathcal{O}(L^3)$.

\section{Wigner $\Delta$ and Recursion}\label{sec:wigrec}
Here we answer the question of how to compute the $\Delta^l_{mn}$ required in the forward 
and backward transforms. We will base our numerical algorithm 
for computation of an arbitrary $\Delta^l_{mn}$ on recursion and exploitation of symmetries.

First we note the symmetry properties (inherited from $d^l_{mn}(\vartheta)$):
\begin{align}
\label{eq:delsym1}
\Delta^l_{-m,n} &= (-1)^{l+n}\Delta^l_{mn}\\
\label{eq:delsym2}
\Delta^l_{m,-n} &= (-1)^{l-m}\Delta^l_{mn}\\
\label{eq:delsym3}
\Delta^l_{mn}   &= (-1)^{n-m}\Delta^l_{nm}.
\end{align}
Using an approach similar to \cite{Trapani:2006he} one can derive an analogous 
Trapani-Navaza (TN) style recursion from Eq.~\eqref{eq:dwigreppol} for half-integer $l,m,n$ values:
\begin{align}
\label{eq:delrectn1}
\Delta^l_{ll} &= \frac{1}{2} \Delta^{l-1}_{l-1,l-1}  \\
\label{eq:delrectn2}
\Delta^l_{ml} &= \sqrt{\frac{l(2l-1)}{2(l+m)(l+m-1)}}\Delta^{l-1}_{m-1,l-1}  \\
\label{eq:delrectn3}
\Delta^l_{mn} &= m\sqrt{\frac{2}{l}} \Delta^l_{m,n+1} & (n=l-1) \\
\label{eq:delrectn4}
\Delta^l_{mn} &= \frac{2m}{\sqrt{(l-n)(l+n+1)}}\Delta^l_{m,n+1}& \\
\nonumber&\hspace*{5em}-\sqrt{\frac{(l-n-1)(l+n+2)}{(l-n)(l+n+1)}}\Delta^l_{m,n+2} & (|n|\leq l-2) \\
\label{eq:delrectn5}
\Delta^l_{l,\frac{1}{2}}&= \sqrt{\frac{2l}{2l+1}}\Delta^{l-1}_{l-1,\frac{1}{2}}
\end{align}
Note that by making use of the symmetries provided by \cref{eq:delsym1,eq:delsym2,eq:delsym3} similar recursion relations 
may be constructed connecting different combinations of subscript indices.

We visualise the possible values of $\Delta^l_{mn}$ up to some maximal band-limit $L$ 
as being arranged in a square pyramidal lattice with $\Delta^l_{mn}$ values 
corresponding to $l=1/2$ and $-1/2\leq m,n\leq 1/2$ occupying the top-most plane; 
$l=3/2$ and $-3/2\leq m,n \leq 3/2$ the next plane down and so forth. 
In a recursive approach one can thus initialise with a single value of $\Delta^{\frac{1}{2}}_{mn}$ and with 
\cref{eq:delrectn1,eq:delrectn2,eq:delrectn5} (together with symmetries) compute those values of $\Delta^l_{mn}$ constrained 
to the surface of the pyramidal lattice. For each fixed $l$ value those $\Delta^l_{mn}$ that occupy the interior of the 
pyramidal structure can be computed using Eq.\eqref{eq:delrectn3} and Eq.\eqref{eq:delrectn4} (together with symmetries).

We note that symmetries allow for a reduction in the total number of elements that must be
calculated explicitly via recursion to $((2l+1)/2)^2$ entries for a fixed $l$ plane and $\frac{1}{12} (2L+3)(2L+1)(L+1)$
total entries for a choice of maximal band-limit $L$. 

In our implementation we initialise with $\Delta^{1/2}_{1/2,-1/2}=1/\sqrt{2}$ and iterate such that for a given $l$ the $m,n$ indices obey the condition $(n<1/2)\wedge(|m|\geq|n|)\wedge(m>-1/2)$.

Working with double precision arithmetic, we have found that the above scheme remains stable up to a band-limit of $L\approx 5173/2$. 
Due to exponential convergence \cite{boyd2001chebyshev} this $L$ will in practical situations be far above the resolution required to accurately sample fields 
for evolution equations and thus is not a concern for this work. 
We note however, that by instead working with ratios such as ${}_r\Delta^l_{mn}=\Delta^l_{mn}/\Delta^l_{m-1,n}$ we have verified that 
it is possible to construct a stable type of hybrid recursion in analogy to \cite{numericalEvolutions2014Beyer}, for $L> 5173/2$.

\section{SWSH consistency checks}\label{sec:consischk}
\begin{figure}[ht]
\centering
\begin{subfigure}[ ]{\includegraphics[width=.49\linewidth] {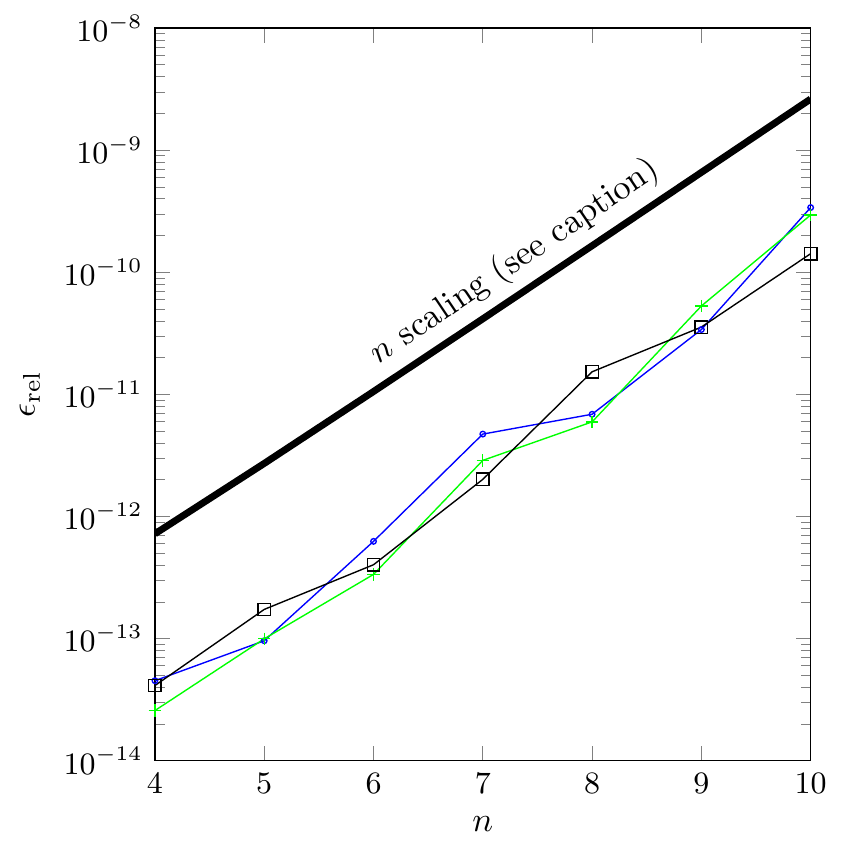}
   \label{fig:fwdbwderr_sub1}
 }%
\end{subfigure}\hfill
\begin{subfigure}[ ]{\includegraphics[width=.49\linewidth]{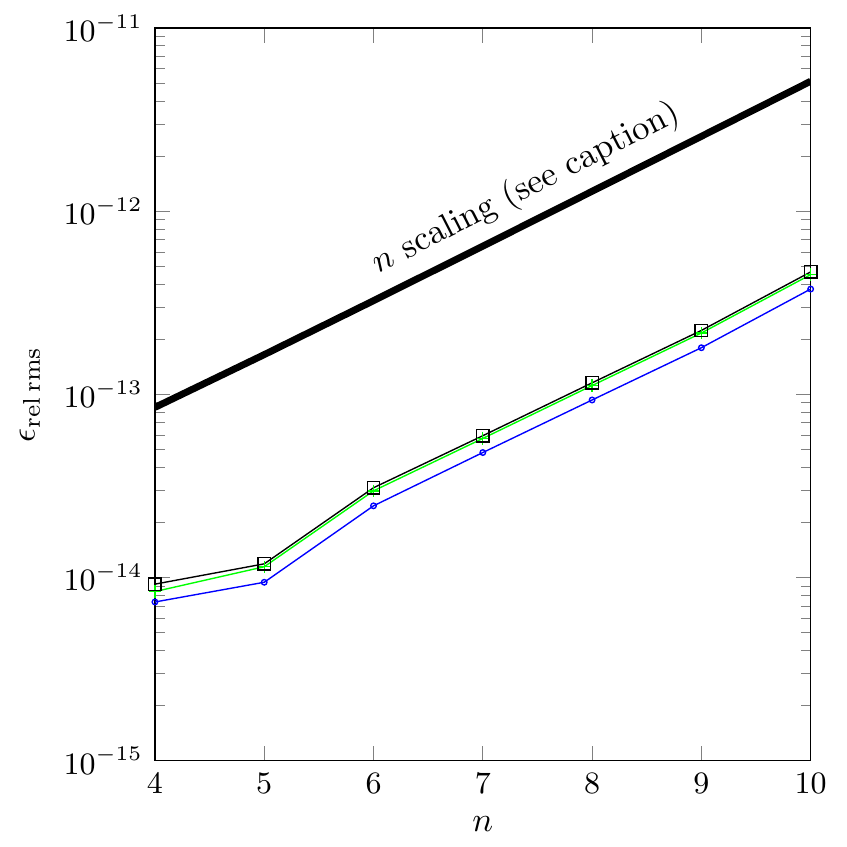}
   \label{fig:fwdbwderr_sub2}
 }%
\end{subfigure}%
\caption{\label{fig:fwdbwderr}
	(Colour online) Averages of numerical error pairs when performing the transform procedure of Eq.~\eqref{eq:fwdbwdtest}. 
	A fixed spin-weight $s$ is chosen and $5$ sets of ${}_sa_{lm}$ are randomly generated as described in the text.
	In both sub-figures the band-limit of transforms is given by $L(n)=(2^n-1)/2$.
	(a) Averaged 
	maximum, absolute relative error $\epsilon_\mathrm{rel}:=\langle\max_{l,m}|1-{}_s\tilde{a}_{lm}/{}_s\tilde{\alpha}_{lm}| \rangle$. 
	The thick black line corresponds to a scaling of $L(n)^2$.
	(b) Averaged 
	maximum, absolute relative rms error 
	$\epsilon_\mathrm{rel\,rms}:=\langle \langle {}_s\tilde{a}_{lm}-{}_s\tilde{\alpha}_{lm} \rangle_\mathrm{rms}/
	\langle {}_s\tilde{a}_{lm} \rangle_\mathrm{rms}\rangle$. 
	The thick black line corresponds to a scaling of $L(n)$. 
	In (a) and (b):
	Green "$+$" corresponds to $s=3/2$;
	blue "$\circ$" corresponds to $s=13/2$;
	black "$\square$" corresponds to $s=-1/2$.
	}
\end{figure}

As a first check on the consistency of the half-integer SWSH algorithm presented above we construct error pairs. Here one populates 
coefficients with random data whereupon a transformation is applied so as to construct the spatial representation of the corresponding function; 
this spatial function is transformed back to coefficients and the associated error may be examined. This procedure may be 
summarised as:
\begin{equation}\label{eq:fwdbwdtest}
  {}_s\tilde{a}_{lm}
  \overset{\mathcal{F}^{-1}}{\longmapsto}
  \;{}_{s}f(\vartheta,\varphi)
  \overset{\mathcal{F}}{\longmapsto}\;{}_{s}\tilde{\alpha}_{lm},
\end{equation}
The real and imaginary parts of ${}_s\tilde{a}_{lm}$ we generate by 
sampling from the continuous uniform random distribution on the interval $[-1,1)$. The numerical error associated with the 
specification of Eq.~\eqref{eq:fwdbwdtest} is shown in Fig.~\ref{fig:fwdbwderr_sub1} and Fig.~\ref{fig:fwdbwderr_sub2}. 
We see that our implementation is indeed consistent, accurate and stable.
Furthermore the scaling $L(n)$ of $\epsilon_\mathrm{rel\,rms}$ (Fig.~\ref{fig:fwdbwderr_sub2})
 for the half-integer spin-weight uniform random data matches the scaling observed for integer spin-weight Gaussian random data observed in 
\cite{Huffenberger:2010hh}.

The product of two spin-weighted functions ${}_{s_1}f$ and ${}_{s_2}g$ on $\mathbb{S}^2$ is a function with spin-weight $s_1+s_2$. If ${}_{s_1}f$ and ${}_{s_2}g$ are comprised of a finite number of SWSH then 
their product is also and thus by selecting a sufficiently large $L$ we exactly sample the resulting product function. 
We can also check the property of exponential convergence; 
to this end define $A_l:=\langle {}_sa_{lm}\rangle_m= \sum_{m}|{}_sa_{lm}|/(2l+1)$, a measure of 
the average magnitude of coefficients at a fixed $l$ value. We expect that given smooth test functions $A_l$ should behave as 
$A_l\sim \alpha \exp(-\kappa l)$ $(\alpha,\kappa\in\mathbb{R})$ for large $l$ \cite{boyd2001chebyshev, katznelson2004harmonic}. 
Smooth half-integer spin-weighted functions may be constructed by 
taking a finite number of SWSH and modulating by the exponential of a smooth spin-weight $0$ function. 
Introduce:
\begin{align}
\label{eq:testFunc1}
{}_{0} g(\vartheta,\varphi) &= \ii\,{}_{0} Y_{3,-1}(\vartheta,\varphi) + 1.1\,{}_{0} Y_{3,1}(\vartheta,\varphi)\\
\label{eq:testFunc2}
{}_{0} \tilde{g}(\vartheta,\varphi) &= \exp\left ( -\,{}_{0} g(\vartheta,\varphi) \right)\\
\label{eq:testFunc3}
{}_{1/2} f(\vartheta,\varphi) &= 1.3\,{}_{1/2}Y_{3/2,1/2}(\vartheta,\varphi)+\ii\,{}_{1/2}Y_{3/2,-1/2}(\vartheta,\varphi)\\
\label{eq:testFunc4}
{}_{1/2} \tilde{f}(\vartheta,\varphi) &= {}_{0} \tilde{g}(\vartheta,\varphi) \, {}_{1/2} f(\vartheta,\varphi)\\
\label{eq:testFunc5}
{}_{3/2} h(\vartheta,\varphi) &= \left[{}_{1/2}Y_{11/2, 3/2}(\vartheta,\varphi)\right]^3\\
\label{eq:testFunc6}
{}_{-1/2} k(\vartheta,\varphi) &= 
0.7i\,{}_{-1/2} Y_{5/2, -1/2}(\vartheta,\varphi)+0.9\,{}_{-1/2} Y_{3/2, 1/2}(\vartheta,\varphi)\\
\label{eq:testFunc7}
{}_{-1/2} \tilde{k}(\vartheta,\varphi) &= {}_{0} \tilde{g}(\vartheta,\varphi) \, {}_{-1/2} k(\vartheta,\varphi)
\end{align}
\begin{figure}[ht]
	\centering
	\includegraphics{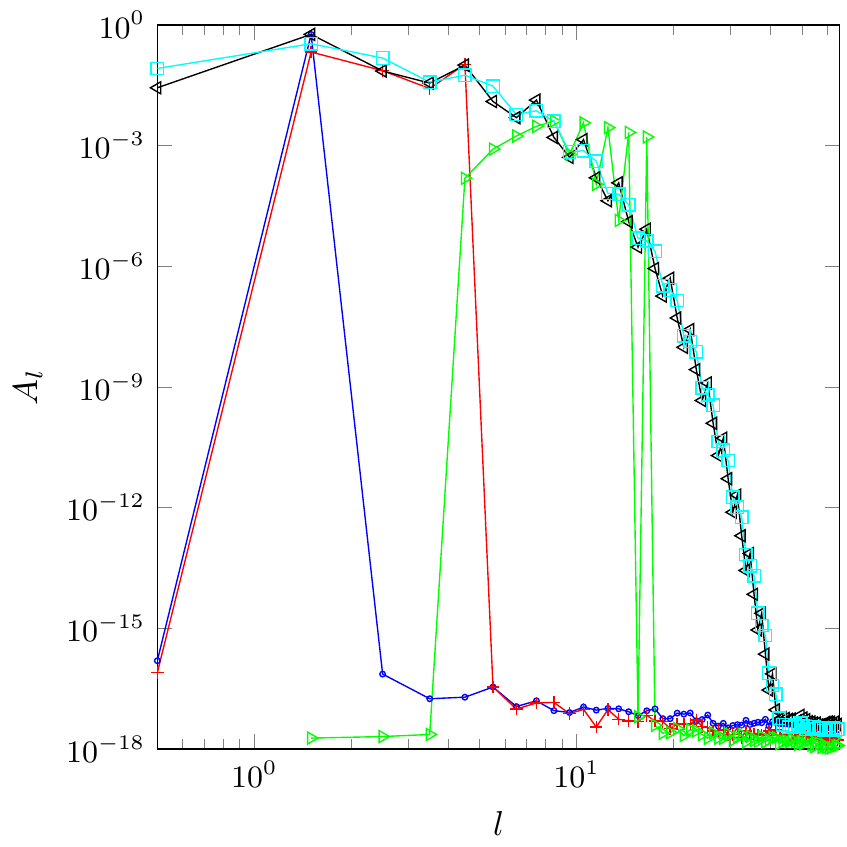}
	\caption{
	(Colour online) The quantity $A_l$ is depicted and corresponds to:
	blue "$\circ$", ${}_{1/2} f(\vartheta,\varphi)$;
	red "$+$", ${}_{1/2} f(\vartheta,\varphi)\,{}_{0} g(\vartheta,\varphi)$;
	black "$\triangleleft$", ${}_{1/2} \tilde{f}(\vartheta,\varphi)$;
	green "$\triangleright$", ${}_{3/2} h(\vartheta,\varphi)$;
	cyan "$\square$, ${}_{-1/2} \tilde{k}(\vartheta,\varphi)$.
	All transforms are performed at a band-limit $L=129/2$ and $s=1/2,\,1/2,\,1/2,\,3/2,\,-1/2$ respectively.
	The functions ${}_{1/2} f(\vartheta,\varphi)$, ${}_{1/2} f(\vartheta,\varphi)\,{}_{0} g(\vartheta,\varphi)$ and 
	${}_{3/2} h(\vartheta,\varphi)$ are completely sampled as the average magnitude of coefficients ($A_l$) falls to numerical 
	round-off for $l$ well below $L$. 
	For ${}_{1/2} \tilde{f}(\vartheta,\varphi)$ and ${}_{-1/2} \tilde{k}(\vartheta,\varphi)$ we observe the expected exponential decay in $A_l$. 
	(See text for details)
	}
	\label{fig:coeffdecay}
\end{figure}
In Fig.~\ref{fig:coeffdecay} we show $A_l$ for various combinations that test the various properties outlined. 
In particular, we observe that functions comprised of a finite number of SWSH are completely captured if $L$ is greater than the highest 
harmonic of the linear combination. 
Products of spin-weighted functions (each individual function comprised of a finite linear combination of SWSH) 
also are exactly resolved to within numerical error for appropriate $L$. 
Furthermore, we observe the crucial property of exponential convergence for smooth functions.

\section{$2+1$ Dirac equation}\label{sec:Dirac}

In order to test the SWSH half-integer algorithm we now numerically solve the
Dirac equation on a 3-dimensional Lorentz manifold $(\mathscr{M},g)$ with
topology $\mathbb{R}\times\mathbb{S}^2$. Since the Dirac equation on a
3-dimensional manifold is less familiar than its 4-dimensional counterpart we
present a detailed derivation in Appendix~\ref{sec:dirac-equation-3} and simply
state the result here.

We consider the manifold $\sM\sim \mathbb{R} \times \mathbb{S}^2$ with the metric
\begin{equation}\label{eq:metrCh}
g=\mathrm{d}t\otimes\mathrm{d}t-\mathcal{F}^{-2}(t,\vartheta,\varphi)
\left(\mathrm{d}\vartheta\otimes\mathrm{d}\vartheta
+\sin^2\vartheta\,\mathrm{d}\varphi\otimes\mathrm{d}\varphi \right),
\end{equation}
where $(\vartheta,\varphi)$ are standard polar coordinates for the 2-sphere. The
function $\mathcal{F}(t,\vartheta,\varphi)$ is a conformal factor relating the
induced metric at every instant of time $t$ on $\mathbb{S}^2$ to the standard metric of
the unit 2-sphere. 

According to Appendix~\ref{sec:dirac-equation-3} the Dirac equation on $\sM$ is
given by the following two equations for a spinor $\psi = [\psi_-,\psi_+]^T$
\begin{align}
  \partial_t \psi_- &= -\ii \mu \psi_- - \frac12 \eth'\F\, \psi_+ - \F\, \eth'
  \psi_+ + \frac{\F_t}\F\,\psi_- ,\label{eq:DiracEOM1}\\
  \partial_t \psi_+ &= \ii \mu \psi_+ - \frac12 \eth\F\, \psi_- - \F\, \eth
  \psi_- + \frac{\F_t}\F\,\psi_+.\label{eq:DiracEOM2}
\end{align}
The spinor components $\psi_\pm$ have spin-weight $\pm\tfrac12$, while
$\mathcal{F}$ has vanishing spin-weight.

\subsection{Pseudo-spectral method}\label{sec:psmethod}
While performing a spectral decomposition of variable coefficient PDEs such as Eq.~\eqref{eq:DiracEOM1} and Eq.~\eqref{eq:DiracEOM2} by assuming functions may be expanded as in Eq.~\eqref{eq:funcs2sphTrunc} and then products decomposed as in Eq.~\eqref{eq:prodSoln} is possible (see \cite{numericalEvolutions2014Beyer} for examples of such expansions in the integer SWSH case) we have found that instead it is far simpler\footnote{ Indeed the pseudo-spectral method generalises more readily to nonlinear problems.} to use a pseudo-spectral (PS) approach.

We outline this as follows: Given the coefficients ${}_{s_1}\tilde{f}_{l_1m_1}$ and 
${}_{s_2}\tilde{g}_{l_2m_2}$ which 
represent the functions ${}_{s_1}f$ and ${}_{s_2}g$ sampled at a band-limit $L$ then the coefficients 
${}_{s_1+s_2}\tilde{a}_{lm}$ corresponding to the associated point-wise product ${}_{s_1}\tilde{f}\cdot{}_{s_2}\tilde{g}$ can be calculated by 
performing the transformations:
\begin{align*}
\mathcal{F}^{-1} :& \left({}_{s_1}\tilde{f}_{l_1m_1}\right)\mapsto {}_{s_1}\tilde{f}, &
\mathcal{F}^{-1} :& \left({}_{s_1}\tilde{g}_{l_1m_1}\right)\mapsto {}_{s_2}\tilde{g},
\end{align*}
subsequently taking the pointwise product and transforming:
\begin{equation*}
\mathcal{F} : {}_{s_1}\tilde{f}\,{}_{s_2}\tilde{g} \mapsto \left( {}_{s_1+s_2}\tilde{a}_{lm} \right),
\end{equation*}
we find an approximation to an expansion utilizing Eq.~\eqref{eq:prodSoln}
directly. We emphasize that this method also easily allows one to take into
account the action of the $\eth,\eth'$ operators on spin-weighted functions by
embedding their action as multiplication (see Eq.~\eqref{eq:ladder1} and
Eq.~\eqref{eq:ladder2}) in coefficient space, together with taking account of
their spin raising and lowering properties when transforms are performed.

We now recast the PDE system of Eq.~\eqref{eq:DiracEOM1} and Eq.~\eqref{eq:DiracEOM2} as an (infinite dimensional) ODE system using the PS method. 
It is convenient to define the auxiliary fields:
\begin{align}
\begin{aligned}\label{eq:auxFields}
\Xi_-:&=\mathcal{F}\eth'\psi_+,&
\Phi_-:&=\frac{1}{2}\psi_+ \eth'\mathcal{F},&
\Psi_-:&=\psi_- \frac{\mathcal{F}_t}{\mathcal{F}}, \\
\Xi_+:&=\mathcal{F}\eth\psi_-,&
\Phi_+:&=\frac{1}{2}\psi_- \eth\mathcal{F},&
\Psi_+:&=\psi_+ \frac{\mathcal{F}_t}{\mathcal{F}},
\end{aligned}
\end{align}
where $\pm$ corresponds to a spin-weight of $\pm1/2$. Using these fields we find that Eq.~\eqref{eq:DiracEOM1} and Eq.~\eqref{eq:DiracEOM2} can 
be written as:
\begin{equation}
\begin{aligned}\label{eq:decompFdep}
\dot{\psi}_{-,lm}(t)&=-\ii\mu\psi_{-,lm}(t)-\Phi_{-,lm}(t)-\Xi_{-,lm}(t)+\Psi_{-,lm}(t)\\
\dot{\psi}_{+,lm}(t)&=\ii\mu\psi_{+,lm}(t)-\Phi_{+,lm}(t)-\Xi_{+,lm}(t)+\Psi_{+,lm}(t)
\end{aligned}
\end{equation}
We note that in the case of $\mathcal{F}=1$, \cref{eq:DiracEOM1,eq:DiracEOM2} yield a 
constant-coefficient PDE system which may be directly decomposed as:
\begin{align}
\begin{aligned}\label{eq:decompF1}
\dot{\psi}_{-,lm}(t)&=-\ii\mu \psi_{-,lm}(t)-\left(l+\frac{1}{2} \right) \psi_{+,lm}(t)\\
\dot{\psi}_{+,lm}(t)&=\ii\mu \psi_{+,lm}(t)+\left(l+\frac{1}{2} \right) \psi_{-,lm}(t)
\end{aligned}
\end{align}
In this case, a solution is readily arrived at:
\begin{equation}\label{eq:analytF1}
\psi_{\pm,lm}(t) = \frac{1}{\omega_l} \left[
\left(\omega_l\cos(\omega_l t)\pm \ii\mu\sin(\omega_l t) \right)\psi_{\pm,lm}(0)
\pm (l+\frac{1}{2})\sin(\omega_l t)\psi_{\mp,lm}(0)
\right],
\end{equation}
where $\omega_l:=\sqrt{\left(l+\frac{1}{2}\right)^2+\mu^2}$.

When $\mathcal{F}\neq 1$ we numerically construct the spatial representation of the function using 
Eq.~\eqref{eq:swshDefn} with spatial sampling to coincide with the half-integer SWSH transformation.
We will also find it useful to introduce the rescaled current component:
\begin{equation}\label{eq:rescCur}
\tilde{j}^0(t,\vartheta,\varphi):=\frac{1}{\mathcal{F}^2(t,\vartheta,\varphi)}j^0(t,\vartheta,\varphi),
\end{equation}
which with the PS method can be computed using the rescaled fields $\tilde{\psi}_\pm:=\psi_\pm/\mathcal{F}^2$.
From Eq.~\eqref{eq:chargeSub} and Eq.~\eqref{eq:rescCur} the probability $Q(\Sigma_\tau)$ can be computed via:
\begin{align}\label{eq:probability}
Q(\Sigma_\tau)=\sum_{l,m} \left(
\overline{\psi}_{-,lm}\tilde{\psi}_{+,lm}+\tilde{\psi}_{-,lm}\overline{\psi}_{+,lm}
\right).
\end{align}

\subsection{Numerical solutions and convergence tests}\label{sec:numericalEOM}

In this section we examine numerical solutions to the EOM \cref{eq:DiracEOM1,eq:DiracEOM2} under three conditions: The case 
$\mathcal{F}=1$ (corresponding to a $\mathbb{S}^2$ spatial geometry), where we numerically 
solve the system of equations given by the decomposition of 
Eq.~\eqref{eq:decompF1}; the case of a static deformation (time-independent) of $\mathcal{F}$; 
a time-dependent $\mathcal{F}$. In the latter two 
cases we work with the decomposition of Eq.~\eqref{eq:decompFdep}.
In each case we are free to select arbitrary initial data for $\psi_\pm$. 
As convergence tests must be performed for any numerical calculation we begin by briefly summarising the procedure we follow 
(based on \cite{numerical2010Baumgarte}) for doing so, subsequently presenting our results. Aside from presenting convergence for the 
fields $\psi_{\pm}$ we also verify that the continuity equation for the current is obeyed and probability is conserved in each case examined.

In using the PS approach of sec.~\ref{sec:psmethod} to compute the time-evolution of $\psi_\pm$ we solve a truncated $(l\leq L)$ ODE system for 
$\psi_{\pm,lm}(t)$. 
Suppose that we use an explicit, temporal integrator with a fixed time-step $\delta t$. The numerically calculated solution for this 
choice of $\delta t$ we denote as $\psi_{\pm,lm}(t;\,\delta t)$. Assume that there exists a Taylor expansion of $\psi_{\pm,lm}(t;\,\delta t)$ 
about the exact solution $\psi_{\pm,lm}(t)$ with error constant terms $\{E_i\}$:
\begin{equation}
\psi_{\pm,lm}(t;\,\delta t) = \psi_{\pm,lm}(t) + \sum_{n=1}^\infty \delta t^n E_n.
\end{equation}
Suppose now that the integrator is of order $p$ so that $\{E_j\}_{j<p}=0$. Thus:
\begin{equation}
\psi_{\pm,lm}(t;\,\delta t)-\psi_{\pm,lm}(t) = \delta t^p E_p+\mathcal{O}(\delta t^{p+1}).
\end{equation}
By successively rescaling $\delta t$ with a constant (here $2$) and comparing $\psi_{\pm,lm}(t;\,\delta t/2^k)$ with $\psi_{\pm,lm}(t)$ we find:
\begin{equation}\label{eq:analytconv}
2^{kp}\left\{\psi_{\pm,lm}\left(t;\,\frac{1}{2^k}\delta t\right) - \psi_{\pm,lm}(t)\right\} 
\rightarrow \delta t^p E_p.
\end{equation}
Equation \eqref{eq:analytconv} may be used in the case when an analytical solution is known. If this is not the case we may instead 
successively compare $\psi_{\pm,lm}(t;\,\delta t/2^k)$ and $\psi_{\pm,lm}(t;\,\delta t/2^{k+1})$ which yields:
\begin{equation}\label{eq:selfconv}
2^{kp}\left\{
\psi_{\pm,lm}\left(t;\,\frac{1}{2^k}\delta t\right) - 
\psi_{\pm,lm}\left(t;\,\frac{1}{2^{k+1}}\delta t\right)
\right\} \rightarrow \left(1-\frac{1}{2^p} \right)\delta t^p E_p=\delta t^p \tilde{E}_p,
\end{equation}
this is known as a self-consistent convergence test (SCCT) \cite{numerical2010Baumgarte}. 
Note that in both cases (Eq.~\eqref{eq:analytconv} and Eq.~\eqref{eq:selfconv}) we 
may also make comparisons of functions in the spatial representation by transforming from coefficient-space. 
In particular, we will consider the maximum, absolute, relative error metric: 
\begin{equation}
\epsilon_r\left(\phi_{\pm}(t,\vartheta,\varphi),\, \psi_{\pm}(t,\vartheta,\varphi)\right)
:=\max_{m,n}\left|1-\frac{\phi_{\pm}(t,m\Delta\vartheta,n\Delta\varphi)}{\psi_{\pm}(t,m\Delta\vartheta,n\Delta\varphi)} \right|,
\end{equation}
which allows for the comparison of the two solutions $\phi_\pm,\, \psi_\pm$ at sampling nodes 
$m\Delta \vartheta$, $n\Delta \varphi$. We measure error in this manner such that 
convergence properties of solutions and any potential instability that may result from the SWSH transformation algorithm may be simultaneously 
inspected.

Having described the methods we use to test convergence of solutions we now
examine some example cases.  We fix the mass-parameter to be $\mu=1.2$ at the
outset and choose a temporal range of $t\in[0,5]$ throughout.  These choices
have been made so as to allow for observation of multiple oscillations of
$\psi_{\pm,lm}(t)$ modes (see Eq.~\eqref{eq:analytF1}) while simultaneously
avoiding exact integer multiple of the frequencies $\omega_l/(2\pi)$.  Note that
all specified initial conditions $\left.\psi_{\pm}\right|_{t=0}$ are constructed
with data as indicated below and then the corresponding coefficients
$\left.\psi_{\pm,lm}\right|_{t=0}$ rescaled by an overall factor so that $Q$ is
normalized to $1$ at $t=0$.  For each numerical evolution the temporal
integrator we choose is the standard, explicit Runge-Kutta $4$th order method
(RK4); thus $p=4$ in Eq.~\eqref{eq:analytconv} and Eq.~\eqref{eq:selfconv}.

\begin{figure}[ht]
	\centering
	\includegraphics{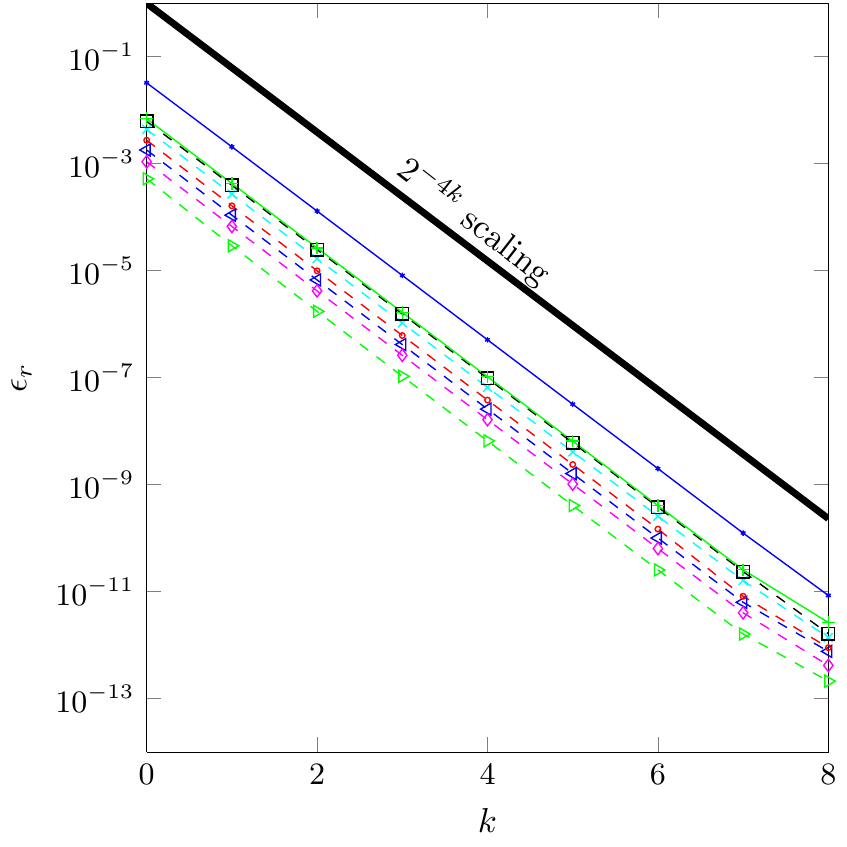}
	\caption{
	(Colour online) Convergence test of numerical solution $\psi_\pm$ (spatial representation) for the case $\mathcal{F}=1$. 
	$\varepsilon_r (\psi_\pm):=\epsilon_r \left(\psi_\pm(t;\delta t^k,\vartheta,\varphi),\,\psi_\pm(t,\vartheta,\varphi) \right)$ is displayed  
	at $t=5$ for a number of steps $N=2^k\times100$. 
	$\psi_\pm(t,\vartheta,\varphi)$ is calculated from Eq.~\eqref{eq:analytF1} and the spatial representation found by the 
	SWSH transformation.
	Blue (dashed), "$\triangleleft$": $L=17/2$, $\varepsilon_r (\psi_-)$;
	Green (dashed), "$\triangleright$": $L=17/2$, $\varepsilon_r (\psi_+)$;
	Red (dashed), "$\circ$": $L=33/2$, $\varepsilon_r (\psi_-)$;
	Cyan (dashed), "$\times$": $L=33/2$, $\varepsilon_r (\psi_+)$;
	Black (dashed), "$\square$": $L=65/2$, $\varepsilon_r (\psi_-)$;
	Magenta (dashed), "$\diamond$": $L=65/2$, $\varepsilon_r (\psi_+)$;
	Blue (solid), "$\ast$": $129/2$, $\varepsilon_r (\psi_-)$;
	Green (solid), "$+$": $129/2$, $\varepsilon_r (\psi_+)$.
	The thick black line corresponds to an expected convergence of $4$th order in time. 
	(See text for discussion)
	}
	\label{fig:eomF1fieldcvgce}
\end{figure}

\begin{figure}[ht]
\centering
\begin{subfigure}[Maximum magnitude of $3$-divergence]{\includegraphics[width=.49\linewidth] {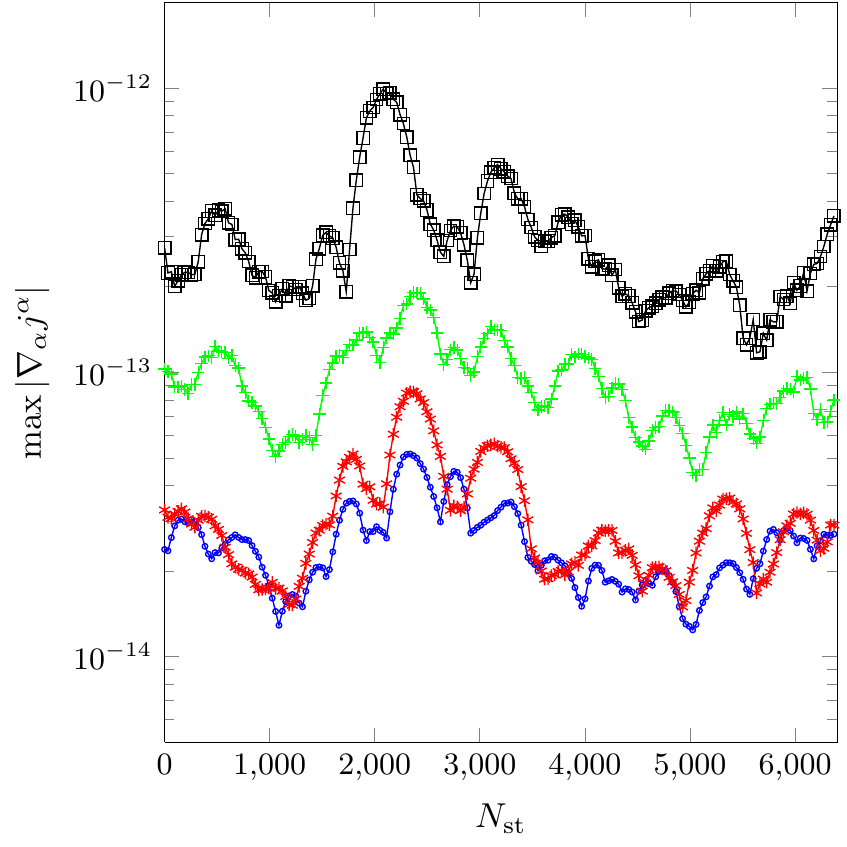}
   \label{fig:contin_S2_sub1}
 }%
\end{subfigure}\hfill
\begin{subfigure}[Probability conservation]{\includegraphics[width=.49\linewidth]{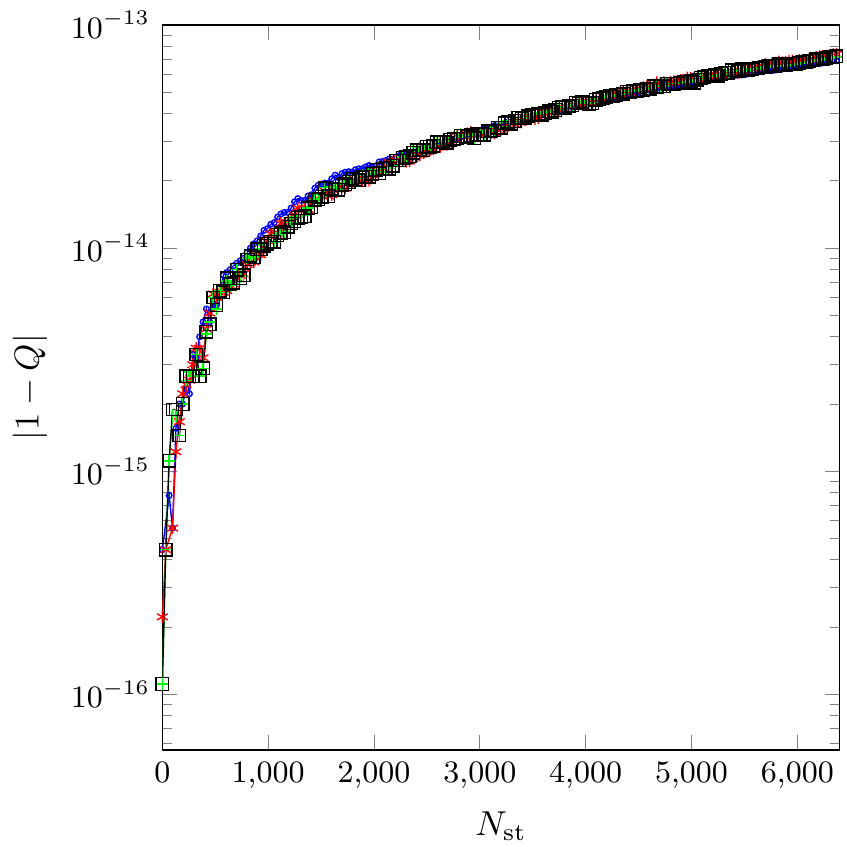}
   \label{fig:prob_S2_sub2}
 }%
\end{subfigure}%
\caption{\label{fig:contin_prob_S2}
	(Colour online) Test of the continuity equation and probability conservation for the case $\mathcal{F}=1$.
	We fix the time-step $\delta t$ to be $5/6400$. $N_\mathrm{st}$ indicates the number of time-steps taken from $t=0$ (to the maximum $t=5$).
	(a) Maximum over all sampled nodes in the spatial representation of $|\nabla_\alpha j^\alpha|$ (based on Eq.~\eqref{eq:divcurspcpt})
	(b) Conservation of probability $Q$ with time (based on Eq.~\eqref{eq:probability})
	In both sub-figures:
	Blue "$\circ$": $L=17/2$;
	Red "$\ast$": $L=33/2$;
	Green "$+$": $L=65/2$;
	Black "$\square$": $L=129/2$. 
	(See text for discussion)
	}
\end{figure}

For $\mathcal{F}=1$ we select the smooth initial data:
\begin{align}
\begin{aligned}\label{eq:eomF1Ini}
\left.\psi_{-}(t,\vartheta,\varphi)\right|_{t=0} &= {}_{-1/2} \tilde{k}(\vartheta,\varphi),
&
\left.\psi_{+}(t,\vartheta,\varphi)\right|_{t=0} &= {}_{1/2} \tilde{f}(\vartheta,\varphi)
\end{aligned}
\end{align}
with ${}_{-1/2} \tilde{k}(\vartheta,\varphi)$ given by Eq.~\eqref{eq:testFunc7} and 
${}_{1/2} \tilde{f}(\vartheta,\varphi)$ given by Eq.~\eqref{eq:testFunc4}.
The system to solve is thus specified by Eq.~\eqref{eq:decompF1} and Eq.~\eqref{eq:eomF1Ini} upon mapping the latter  
$\left.\psi_\pm(t,\vartheta,\varphi)\right|_{t=0}$ to coefficient space using the SWSH transformation algorithm. 
The results of convergence tests (based on Eq.~\eqref{eq:analytconv} with the analytical solution of Eq.~\eqref{eq:analytF1}) 
of our numerical solution are shown in Fig.~\ref{fig:eomF1fieldcvgce}. We find excellent agreement with the expected $4$th order convergence 
in time for initial data and parameters specified. In particular, we see that the numerical error of the solution is dominated by the temporal 
discretization.
In Fig.~\ref{fig:contin_S2_sub1} and Fig.~\ref{fig:prob_S2_sub2} we provide representative examples 
of the maximum absolute value of the $3$-divergence $\nabla_\alpha j^\alpha$ (expected to be $0$ by Eq.~\eqref{eq:divcurspcpt}) 
and probability conservation $|1-Q|$ (again expected to be $0$) at a fixed number of time-steps (temporal discretization). 
This allows for the examination of the effect of varying the band-limit $L$ on the evolution of the 
smooth initial data of Eq.~\eqref{eq:eomF1Ini}.
In Fig.~\ref{fig:contin_S2_sub1} we observe that the numerical error is greatest for $L=129/2$ and minimizes for $L=17/2$ -- the reason for this 
can be explained as follows:
From the evolution equation (Eq.~\eqref{eq:decompF1}) we see that coupling does not occur between distinct $l,m$ modes, only between fixed 
$l,m$ of $\psi_{\pm,lm}$. Hence we see that if for given $l,m$ it is the case that 
$\left.\psi_{-,lm}\right|_{t=0}=\left.\psi_{+,lm}\right|_{t=0}=0$ then to within 
numerical tolerance these coefficients should remain $0$ over the course of the evolution. 
Thus for a finite $L$ we effectively provide a choice of initial data which can be viewed as exact at the specified $L$. The error arising from 
the spectral components of $\psi_\pm$ can hence be entirely attributed to that of the SWSH transformation itself, 
which scales with increasing $L$ as shown in Fig.~\ref{fig:fwdbwderr}.
This behaviour can also be observed in Fig.~\ref{fig:eomF1fieldcvgce} by comparing the numerical error associated with components 
$\psi_\pm$ at a fixed $k$ for differing $L$. 
In Fig.~\ref{fig:contin_prob_S2} we examine the conservation of probability with the quantity $|1-Q|$ where $Q$ is calculated according to 
Eq.~\eqref{eq:probability} and as $\mathcal{F}=1$ we may avoid rescaling to $\tilde{\psi}_\pm$ as is required in more general cases. As no 
SWSH transformations are required to compute this quantity once $\psi_{\pm,lm}(t)$ is known 
we find that upon changing $L$ the previous associated numerical error is not accumulated and the results for $|1-Q|$ 
at differing $L$ coincide.

\begin{figure}[ht]
	\centering
	\includegraphics{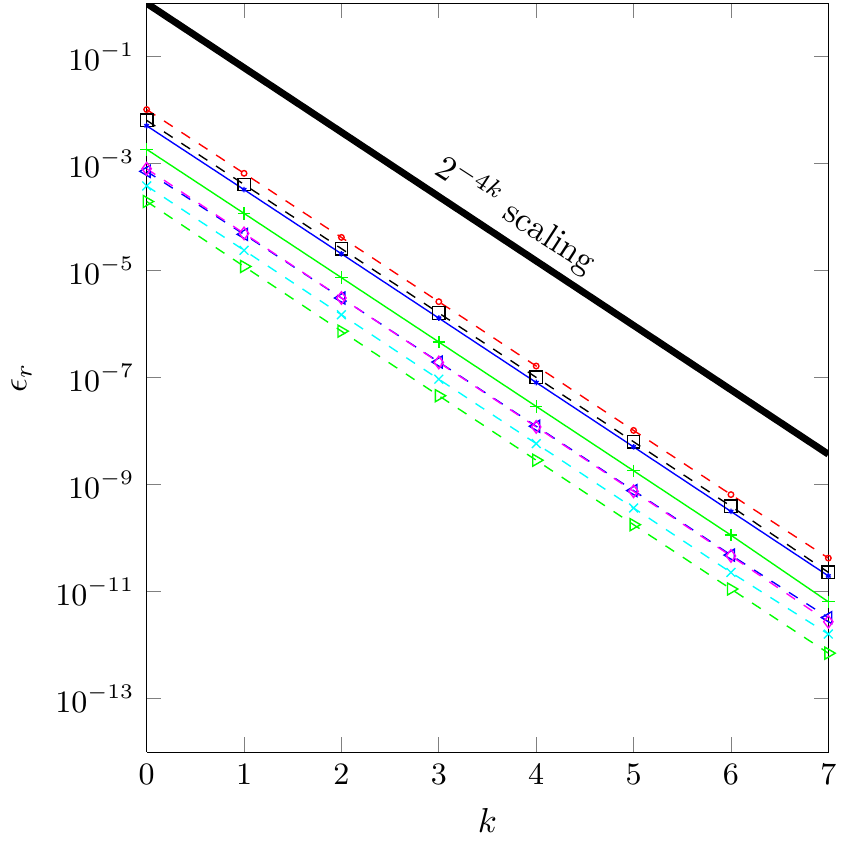}
	\caption{
	(Colour online) SCCT of numerical solution $\psi_\pm$ (spatial representation) for the case of 
	static $\mathcal{F}$ (Eq.~\eqref{eq:eomFdef}). 
	$\varepsilon_r (\psi_\pm):=\epsilon_r \left(\psi_\pm(t;\delta t^k,\vartheta,\varphi),
	\,\psi_\pm(t;\delta t^{k+1},\vartheta,\varphi) \right)$ is displayed  
	at $t=5$ for a number of steps $N=2^k\times100$. 
	Blue (dashed), "$\triangleleft$": $L=17/2$, $\varepsilon_r (\psi_-)$;
	Green (dashed), "$\triangleright$": $L=17/2$, $\varepsilon_r (\psi_+)$;
	Red (dashed), "$\circ$": $L=33/2$, $\varepsilon_r (\psi_-)$;
	Cyan (dashed), "$\times$": $L=33/2$, $\varepsilon_r (\psi_+)$;
	Black (dashed), "$\square$": $L=65/2$, $\varepsilon_r (\psi_-)$;
	Magenta (dashed), "$\diamond$": $L=65/2$, $\varepsilon_r (\psi_+)$;
	Blue (solid), "$\ast$": $129/2$, $\varepsilon_r (\psi_-)$;
	Green (solid), "$+$": $129/2$, $\varepsilon_r (\psi_+)$.
	The thick black line corresponds to an expected convergence of $4$th order in time. 
	(See text for discussion)
	}
	\label{fig:eomFdiffieldcvgce}
\end{figure}

\begin{figure}[ht]
\centering
\begin{subfigure}[Maximum magnitude of $3$-divergence]{\includegraphics[width=.49\linewidth] {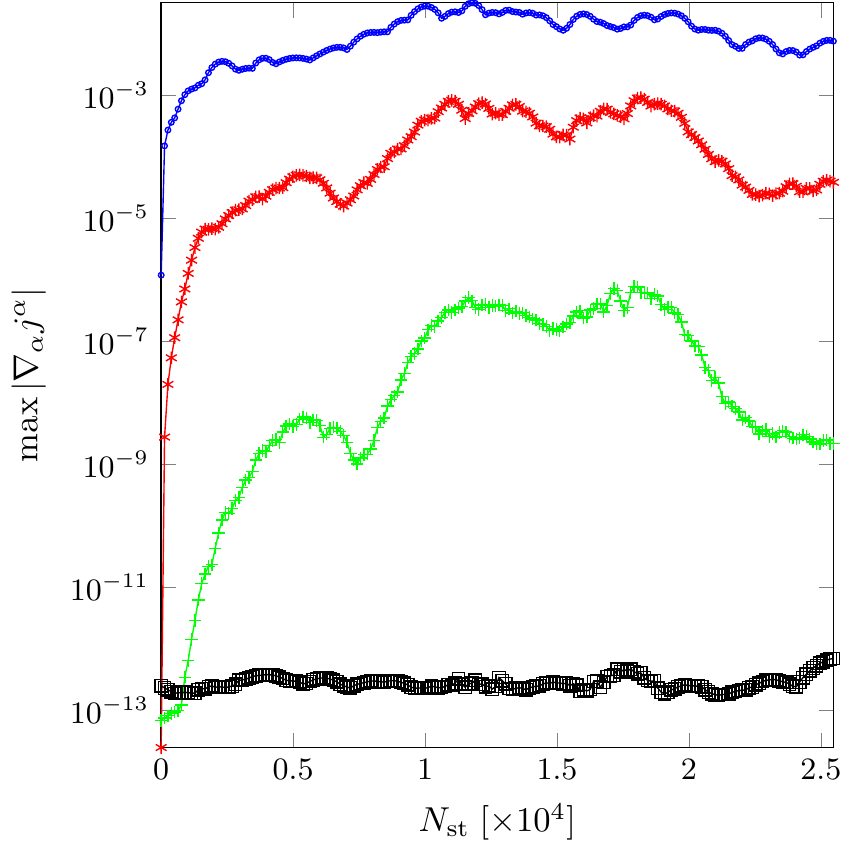}
   \label{fig:contin_diffeo_sub1}
 }%
\end{subfigure}\hfill
\begin{subfigure}[Probability conservation]{\includegraphics[width=.49\linewidth]{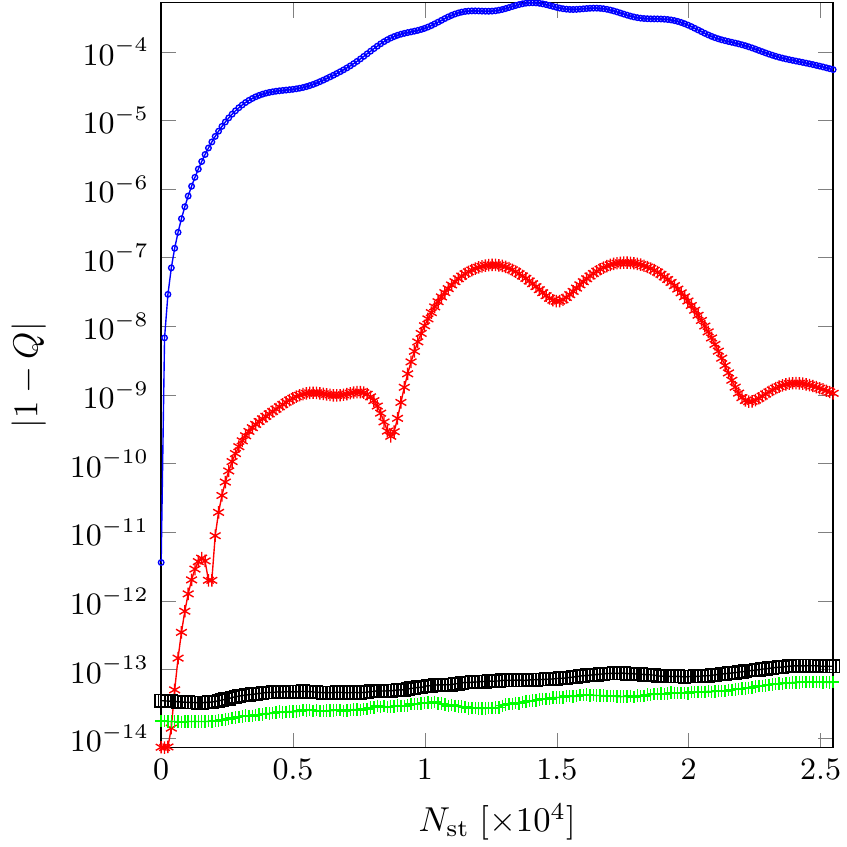}
   \label{fig:prob_diffeo_sub2}
 }%
\end{subfigure}%
\caption{\label{fig:contin_prob_diffeo} (Colour online) Test of the continuity
  equation and probability conservation for the case of static $\mathcal{F}$
  (Eq.~\eqref{eq:eomFdef}).  We fix the time-step $\delta t$ to be
  $5/6400$. $N_\mathrm{st}$ indicates the number of time-steps taken from $t=0$
  (to the maximum $t=5$).  (a) Maximum over all sampled nodes in the spatial
  representation of $|\nabla_\alpha j^\alpha|$ (based on
  Eq.~\eqref{eq:divcurspcpt}) (b) Conservation of probability $Q$ with time
  (based on Eq.~\eqref{eq:probability}) In both sub-figures: Blue "$\circ$":
  $L=17/2$; Red "$\ast$": $L=33/2$; Green "$+$": $L=65/2$; Black "$\square$":
  $L=129/2$.  (See text for discussion) }
\end{figure}

\begin{figure}[ht]
	\centering
	\includegraphics{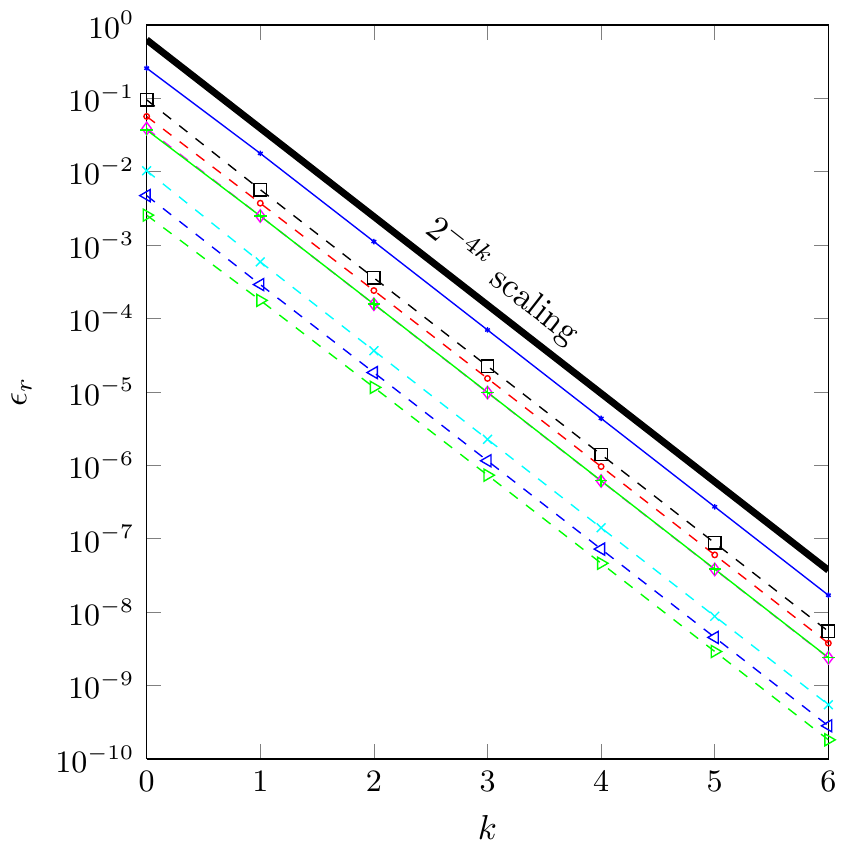}
	\caption{
	(Colour online) SCCT of numerical solution $\psi_\pm$ (spatial representation) for the case of 
	time-dependent $\mathcal{F}$ (Eq.~\eqref{eq:Ftdepg}). 
	$\varepsilon_r (\psi_\pm):=\epsilon_r \left(\psi_\pm(t;\delta t^k,\vartheta,\varphi),
	\,\psi_\pm(t;\delta t^{k+1},\vartheta,\varphi) \right)$ is displayed  
	at $t=5$ for a number of steps $N=2^k\times200$. 
	Blue (dashed), "$\triangleleft$": $L=17/2$, $\varepsilon_r (\psi_-)$;
	Green (dashed), "$\triangleright$": $L=17/2$, $\varepsilon_r (\psi_+)$;
	Red (dashed), "$\circ$": $L=33/2$, $\varepsilon_r (\psi_-)$;
	Cyan (dashed), "$\times$": $L=33/2$, $\varepsilon_r (\psi_+)$;
	Black (dashed), "$\square$": $L=65/2$, $\varepsilon_r (\psi_-)$;
	Magenta (dashed), "$\diamond$": $L=65/2$, $\varepsilon_r (\psi_+)$;
	Blue (solid), "$\ast$": $129/2$, $\varepsilon_r (\psi_-)$;
	Green (solid), "$+$": $129/2$, $\varepsilon_r (\psi_+)$.
	The thick black line corresponds to an expected convergence of $4$th order in time. 
	(See text for discussion)
	}
	\label{fig:eomFtdepfieldcvgce}
\end{figure}

\begin{figure}[ht]
\centering
\begin{subfigure}[Maximum magnitude of $3$-divergence]{\includegraphics[width=.49\linewidth] {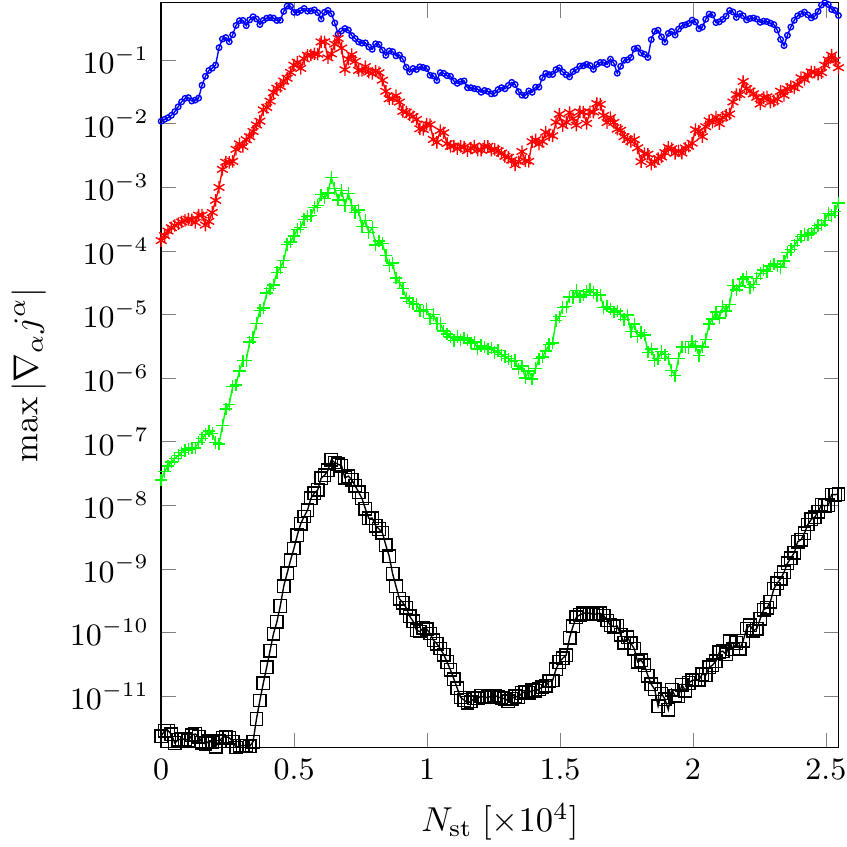}
   \label{fig:contin_tdep_sub1}
 }%
\end{subfigure}\hfill
\begin{subfigure}[Probability conservation]{\includegraphics[width=.49\linewidth]{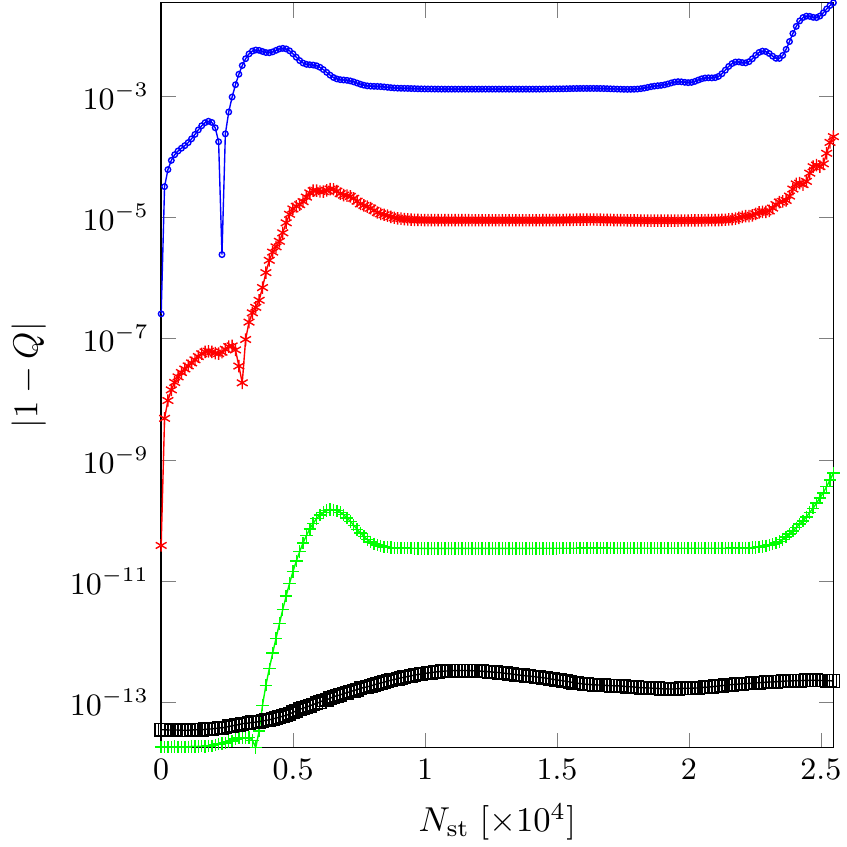}
   \label{fig:prob_tdep_sub2}
 }%
\end{subfigure}%
\caption{\label{fig:contin_prob_tdep}
	(Colour online) Test of the continuity equation and probability conservation for the case of 
	time-dependent $\mathcal{F}$ (Eq.~\eqref{eq:Ftdepg}).
	We fix the time-step $\delta t$ to be $5/6400$. $N_\mathrm{st}$ indicates the number of time-steps taken from $t=0$ (to the maximum $t=5$).
	(a) Maximum over all sampled nodes in the spatial representation of $|\nabla_\alpha j^\alpha|$ (based on Eq.~\eqref{eq:divcurspcpt})
	(b) Conservation of probability $Q$ with time (based on Eq.~\eqref{eq:probability})
	In both sub-figures:
	Blue "$\circ$": $L=17/2$;
	Red "$\ast$": $L=33/2$;
	Green "$+$": $L=65/2$;
	Black "$\square$": $L=129/2$. 
	(See text for discussion)
		}
\end{figure}

For a time independent $\mathcal{F}$ we select the (finite $L$) initial data $\left.\psi_{\pm,lm}(t)\right|_{t=0}$:
\begin{equation}
\label{eq:eomFdefIniA}
\begin{gathered}
\psi_-{}_{\frac{1}{2},-\frac{1}{2}} = 2, \quad
\psi_-{}_{\frac{1}{2},\frac{1}{2}} =  
\psi_-{}_{\frac{3}{2},-\frac{3}{2}} = 
\psi_-{}_{\frac{3}{2},\frac{3}{2}} = 1, \\
\psi_-{}_{\frac{7}{2},-\frac{5}{2}} = \ii, \quad
\psi_-{}_{\frac{7}{2},\frac{1}{2}} = -2\ii, \quad
\psi_-{}_{\frac{7}{2},\frac{7}{2}} = 1,
\end{gathered}
\end{equation}
\begin{equation}\label{eq:eomFdefIniB}
\begin{aligned}
\psi_+{}_{\frac{1}{2},-\frac{1}{2}} &= 3, &
\psi_+{}_{\frac{1}{2},\frac{1}{2}} = & 
\psi_+{}_{\frac{3}{2},-\frac{3}{2}} = 1, &
\psi_+{}_{\frac{3}{2},\frac{3}{2}} &= 5, &
\psi_+{}_{\frac{7}{2},-\frac{7}{2}} &= 3, \\
\psi_+{}_{\frac{7}{2},-\frac{5}{2}} &= 2, &
\psi_+{}_{\frac{7}{2},-\frac{1}{2}} &= 1, &
\psi_+{}_{\frac{7}{2},\frac{3}{2}} &= 2, &
\psi_+{}_{\frac{7}{2},\frac{5}{2}} &= 1, 
\end{aligned}
\end{equation}
and $\mathcal{F}$ we choose as:
\begin{align}\label{eq:eomFdef}
\mathcal{F}_{0,0}&=4, &
\mathcal{F}_{1,-1}&=-\mathcal{F}_{1,1}=\frac{1}{2}, &
\mathcal{F}_{4,2}&=\mathcal{F}_{4,-2}=\frac{1}{10}.
\end{align}
Together with the above specifications we make use of auxiliary fields as defined in 
Eq.~\eqref{eq:auxFields} with $\Psi_\pm(t)=0\Longrightarrow \Psi_{\pm,lm}(t)=0$ and the EOM decomposition of Eq.~\eqref{eq:decompFdep}.
In this case instead of comparing with an analytical solution we perform a self-consistent convergence based on Eq.~\eqref{eq:selfconv}. 
The result of this is presented in Fig.~\ref{fig:eomFdiffieldcvgce}.
We observe the expected $4$th order convergence in time for the numerical
solution $\psi_{\pm}(t,\vartheta,\varphi)$ at the band limits tested.  In
Fig.~\ref{fig:contin_diffeo_sub1} and Fig.~\ref{fig:prob_diffeo_sub2} we show
representative examples of how the numerical solution obeys the continuity
equation and probability conservation respectively\footnote{The amount of raw
  data that can be generated at higher $L$ can grow dramatically - during the
  calculation we thus only retain the coefficients $\psi_{\pm,lm}(t)$ at evenly
  interspersed points on the temporal grid as presented in figures.}. Note that
in order to calculate $Q$ for this case we rescale fields as in
Eq.~\eqref{eq:probability} and use the full prescription of the PS method (see
sec.~\ref{sec:psmethod}).  In contrast to the $\mathcal{F}=1$ case we now have a
variable coefficient EOM and coupling between distinct $l,m$ modes of
$\psi_{\pm,lm}$ occurs, thus sampling at a sufficiently high band-limit $L$ in
addition to choosing a sufficiently fine temporal grid is required in order to
resolve conservation properties accurately.
Finally we consider a time-dependent $\mathcal{F}$. The selection we make is linear interpolation in time between 
initial $g(\vartheta,\varphi)$ and final $h(\vartheta,\varphi)$ static deformations of $\mathbb{S}^2$.
That is, $\mathcal{F}(t,\vartheta,\varphi):=(1-t/t_f)g(\vartheta,\varphi)+(t/t_f) h(\vartheta,\varphi)$ and $t_f=5$.
The (real) functions $g$ and $h$ we choose to have non-zero $g_{l,m}$ and $h_{l,m}$ coefficients:
\begin{align}
\begin{aligned}\label{eq:Ftdepg}
g_{0,0} &= 8, &
g_{1,-1} &=-g_{1,1}=\frac{5}{2},& 
g_{4,-2} =g_{4,2}=\frac{1}{10} \\
h_{0,0} &= 8, &
h_{1,-1} &=-h_{1,1}=-\frac{5}{2},& 
h_{4,-2} =h_{4,2}=\frac{1}{10} \\
\end{aligned}
\end{align}
Initial data $\left.\psi_{\pm,lm}(t)\right|_{t=0}$ is selected as in Eq.~\eqref{eq:eomFdefIniA} and Eq.~\eqref{eq:eomFdefIniB}.
Together with the above specifications we again make use of auxiliary fields as defined in 
Eq.~\eqref{eq:auxFields} (note however that $\Psi_\pm(t)\neq0$) and the EOM decomposition of Eq.~\eqref{eq:decompFdep}.
We again perform a self-consistent convergence based on Eq.~\eqref{eq:selfconv}. 
The result of this is presented in Fig.~\ref{fig:eomFtdepfieldcvgce}.
Once again, excellent agreement with the expected $4$th order convergence is
observed with respect to the temporal scheme.  We are again in a situation with
a variable coefficient EOM, however now it also becomes non-autonomous. The
continuity equation, together with probability conservation must be obeyed in
this case also, and indeed we find that provided a sufficiently high band-limit
$L$ is chosen we may resolve these stated properties in our numerical solution
(See Fig.~\ref{fig:contin_tdep_sub1} and Fig.~\ref{fig:prob_tdep_sub2}).

\subsection{Dirac equation: collapsing background geometry}
Having analysed the numerical properties of our implementation of a
pseudo-spectral method for the Dirac equation we now turn to potential
applications.  We consider a $(2+1)$ dimensional analogue of the usual
Friedmann-Robertson-Walker (FRW) space-time in comoving coordinates. The
particular physical scenario we wish to model is an imploding universe with
pressure $P$ and density $\rho$ equal to zero.  We take the scale factor to be
$a(t)=1-t=\mathcal{F}(t)^{-1}$. Thus, we see that at $t=0$ the background
spatial geometry coincides with that of $\mathbb{S}^2$ upon which initial data
for the Dirac equation must be provided.

\begin{figure}[ht]
	\centering
	\includegraphics{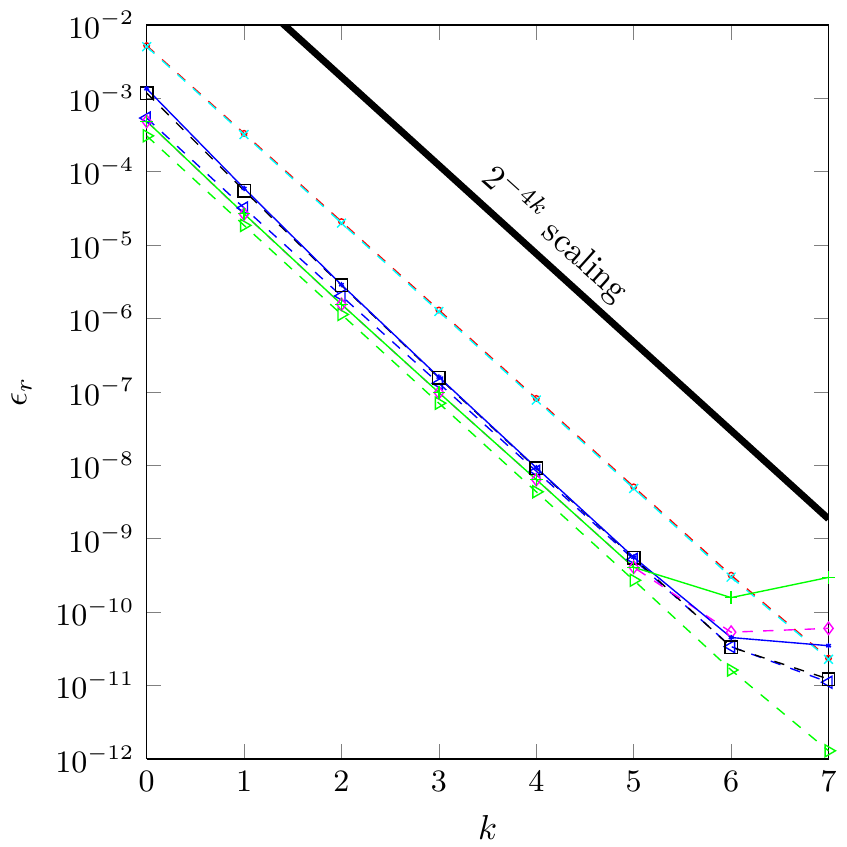}
	\caption{
	(Colour online) SCCT of numerical solution $\psi_\pm$ (spatial representation) for the case of 
	$\mathcal{F}(t)^{-1}=a(t)=1-t$. 
	$\varepsilon_r (\psi_\pm):=\epsilon_r \left(\psi_\pm(t;\delta t^k,\vartheta,\varphi),
	\,\psi_\pm(t;\delta t^{k+1},\vartheta,\varphi) \right)$ is displayed  
	at $t=0.99$ for a number of steps $N=2^k\times4000$. 
	Blue (dashed), "$\triangleleft$": $L=17/2$, $\varepsilon_r (\psi_-)$;
	Green (dashed), "$\triangleright$": $L=17/2$, $\varepsilon_r (\psi_+)$;
	Red (dashed), "$\circ$": $L=33/2$, $\varepsilon_r (\psi_-)$;
	Cyan (dashed), "$\times$": $L=33/2$, $\varepsilon_r (\psi_+)$;
	Black (dashed), "$\square$": $L=65/2$, $\varepsilon_r (\psi_-)$;
	Magenta (dashed), "$\diamond$": $L=65/2$, $\varepsilon_r (\psi_+)$;
	Blue (solid), "$\ast$": $129/2$, $\varepsilon_r (\psi_-)$;
	Green (solid), "$+$": $129/2$, $\varepsilon_r (\psi_+)$.
	The thick black line corresponds to an expected convergence of $4$th order in time. 
	(See text for discussion)
	}
	\label{fig:eomFRWfieldcvgce}
\end{figure}

\begin{figure}[ht]
\centering
\begin{subfigure}[Maximum magnitude of $3$-divergence]{\includegraphics[width=.49\linewidth] {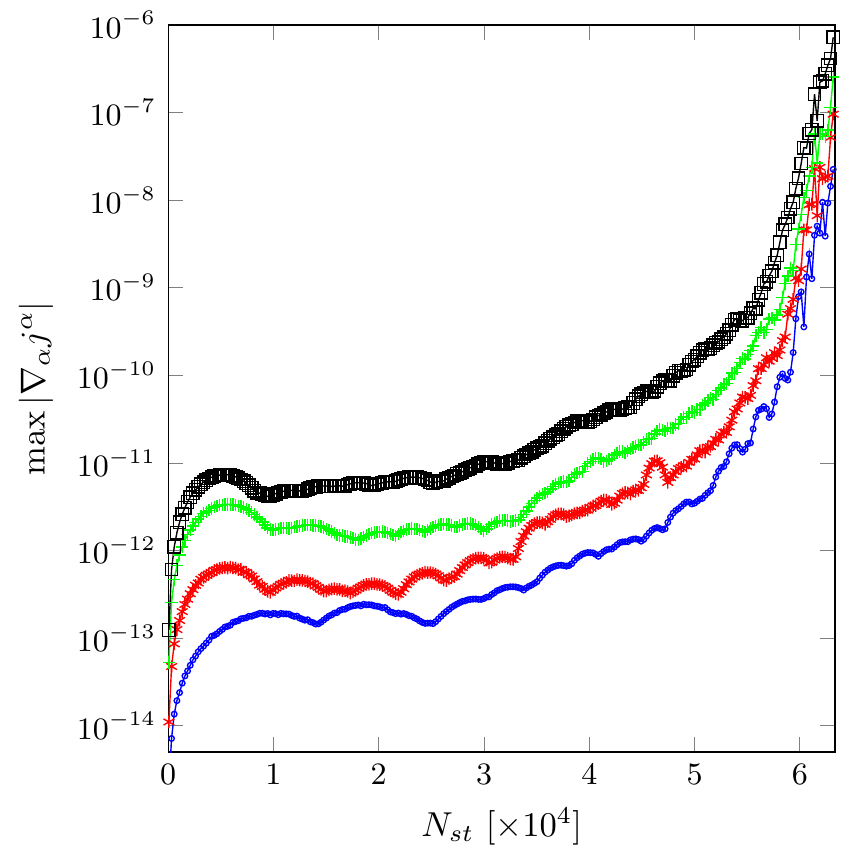}
   \label{fig:contin_frw_sub1}
 }%
\end{subfigure}\hfill
\begin{subfigure}[Probability conservation]{\includegraphics[width=.49\linewidth]{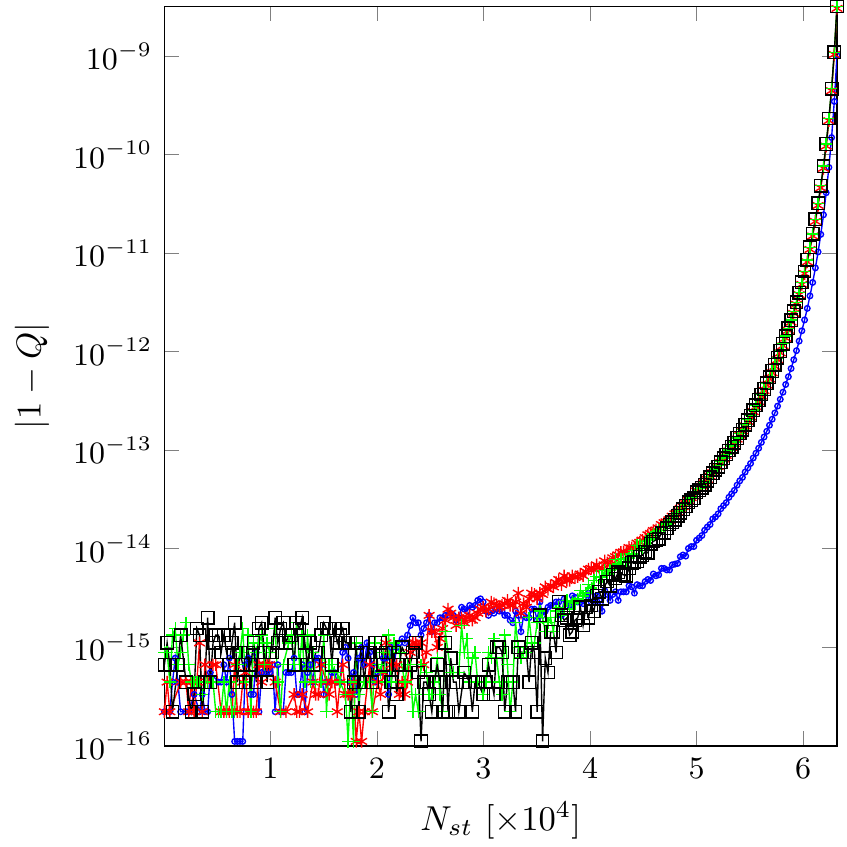}
   \label{fig:prob_frw_sub2}
 }%
\end{subfigure}%
\caption{\label{fig:contin_prob_tdep_2}
	(Colour online) Test of the continuity equation and probability conservation for the case of 
	$\mathcal{F}(t)^{-1}=a(t)=1-t$.
	We fix the time-step $\delta t$ to be $0.99/6.4\times10^{4}$. 
	$N_\mathrm{st}$ indicates the number of time-steps taken from $t=0$ (to the maximum $t=0.99$).
	(a) Maximum over all sampled nodes in the spatial representation of $|\nabla_\alpha j^\alpha|$.
	(b) Conservation of probability $Q$ with time. 
	In both subfigures:
	Blue "$\circ$": $L=17/2$;
	Red "$\ast$": $L=33/2$;
	Green "$+$": $L=65/2$;
	Black "$\square$": $L=129/2$. 
	(See text for discussion)
		}
\end{figure}

\begin{figure}[ht]
\centering
\begin{subfigure}[$t=0$]{\label{fig:DiracFRWSoln_a}
    \includegraphics[width=.47\linewidth] {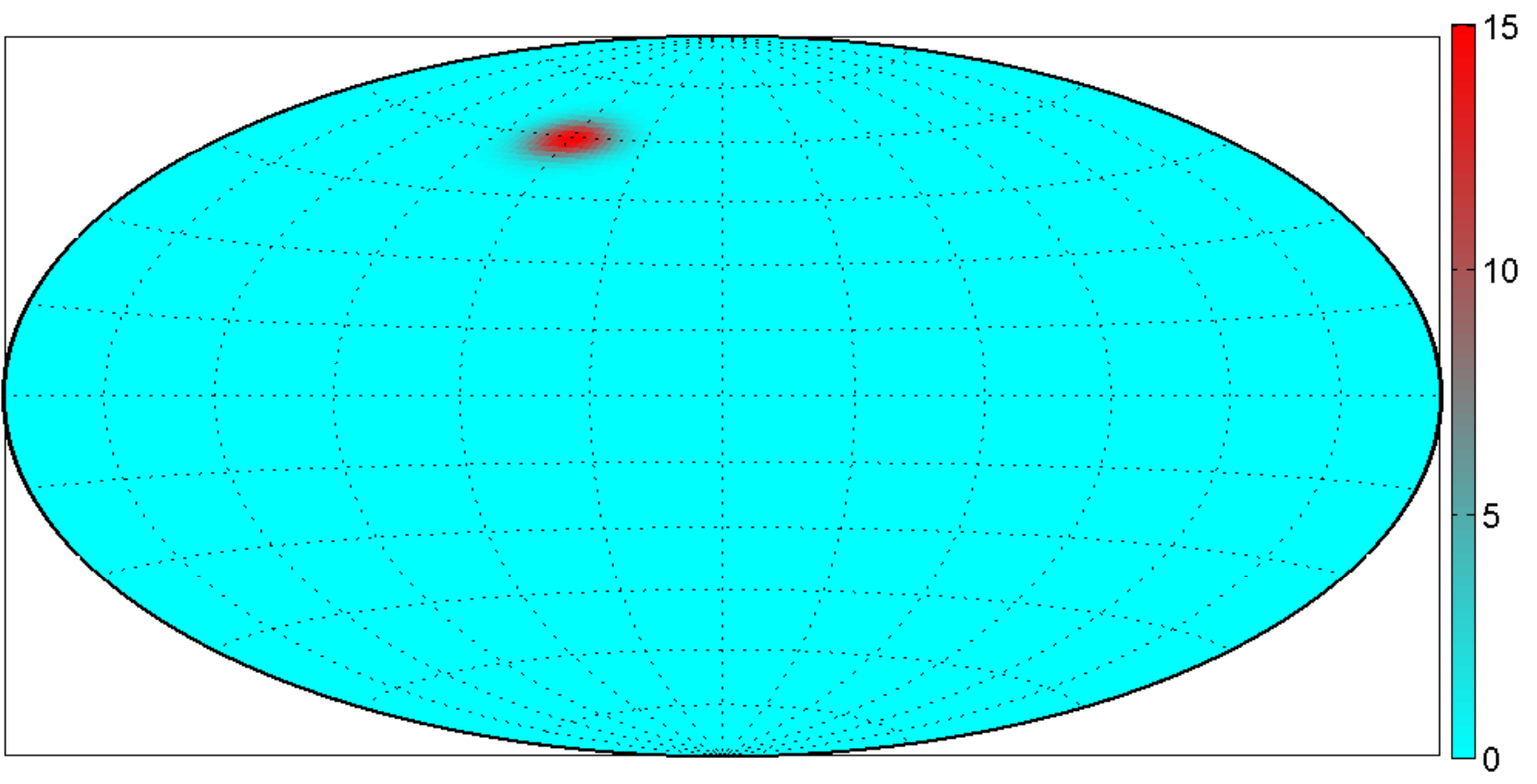}
    }
\end{subfigure}
\hfill
 \begin{subfigure}[$t=0.2477$]{\label{fig:DiracFRWSoln_b}
     \includegraphics[width=.47\linewidth]{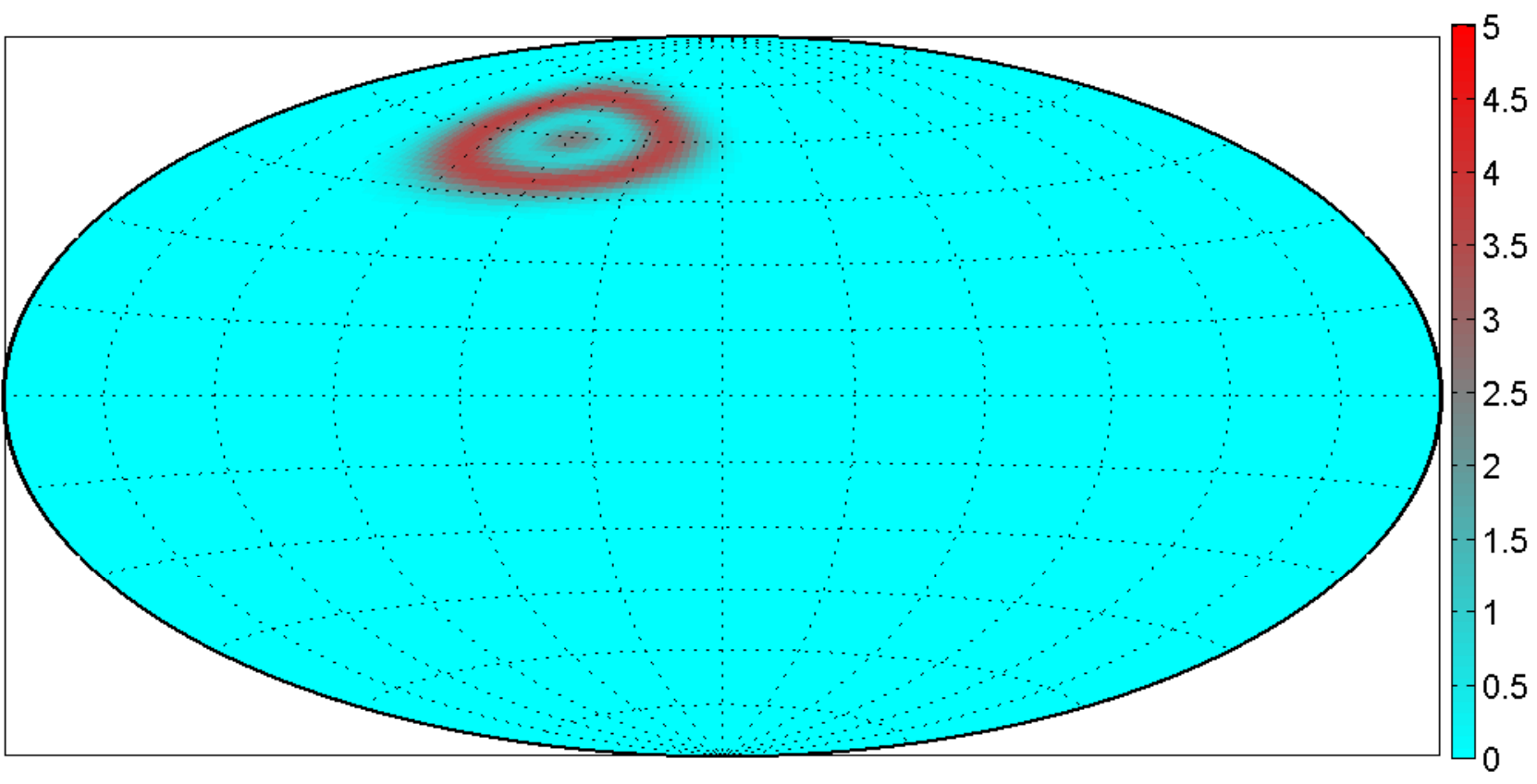}
     }
\end{subfigure}
\begin{subfigure}[$t=0.7432$]{\label{fig:DiracFRWSoln_c}
    \includegraphics[width=.47\linewidth] {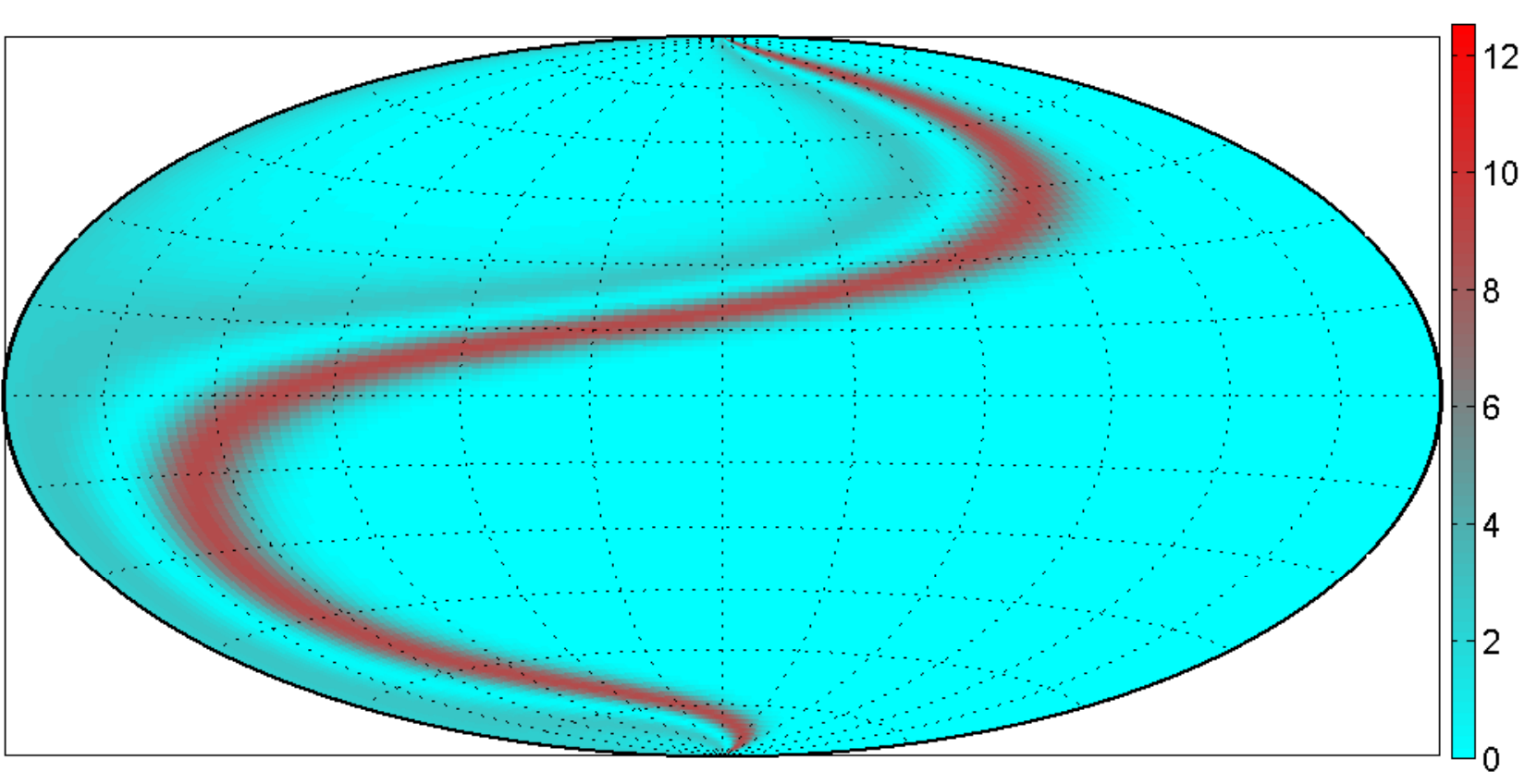}
    }
\end{subfigure}
\hfill
 \begin{subfigure}[$t=0.9414$]{\label{fig:DiracFRWSoln_d}
     \includegraphics[width=.47\linewidth]{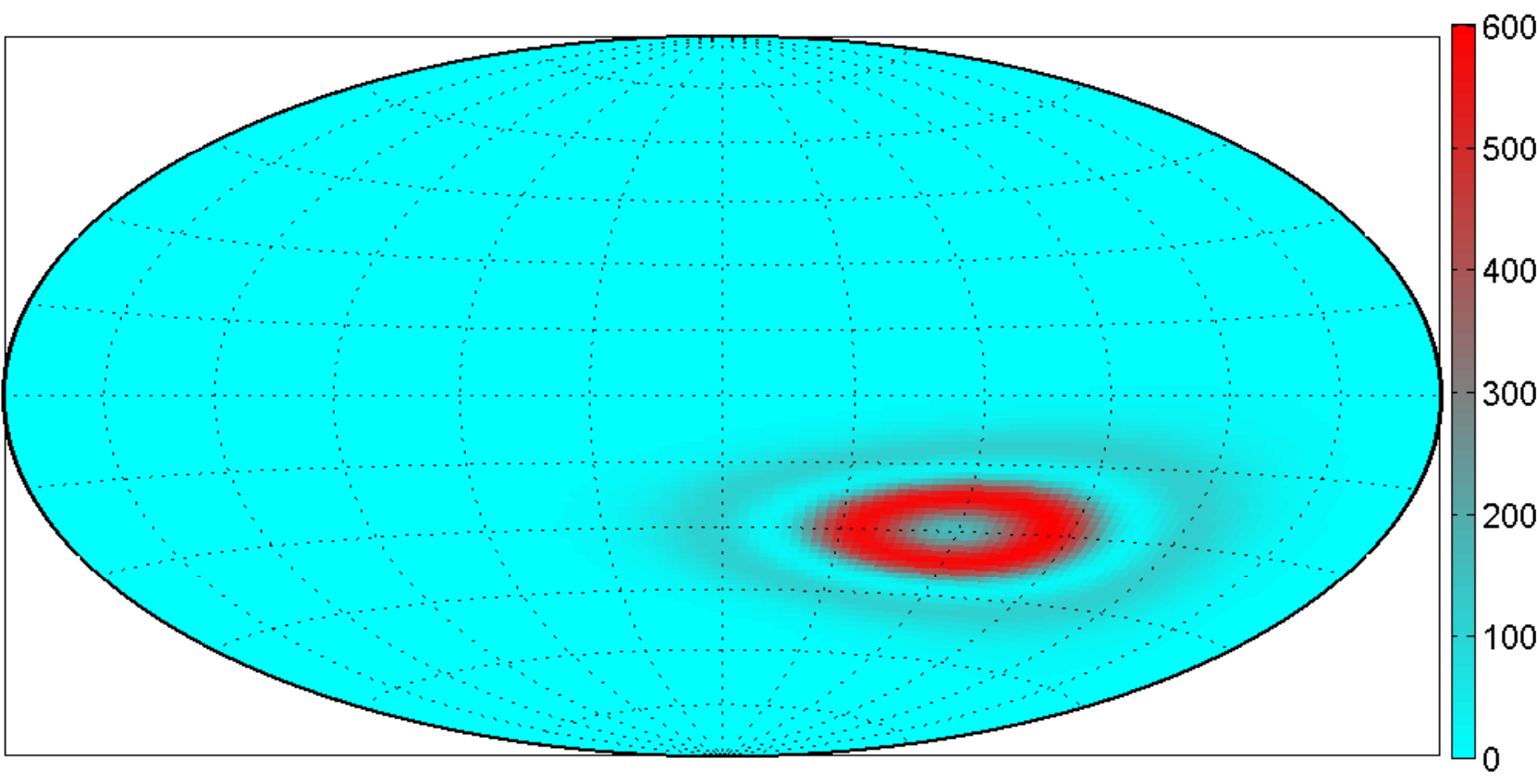}
     }
\end{subfigure}
\caption{
\label{fig:DiracFRWSoln}
(Colour online) Aitoff-Hammer projection of the probability density $j^0$
associated with the $(2+1)$ dimensional Dirac equation on FRW background.  Note
that during the course of the time evolution the spatial radius $a(t)$ of the
spherical projection in each sub-figure is distinct, furthermore the colour
scaling also varies. See text for discussion.  }
\end{figure}

For many practical purposes it is sufficient to consider an initial Gaussian
state when discussing the dynamics of the Dirac equation. For a $(n-1)$-sphere
embedded in $\mathbb{R}^n$ the von Mises-Fisher distribution serves as an
analogue to the planar Gaussian distribution
\cite{Mardia2009DirectionalStatistics}.  The probability density function is
given by:
\begin{equation*}
f_n(\mathbf{x};\,\mathbf{x}_0,\,\kappa)=\frac{\kappa^{n/2-1}}{(2\pi)^{n/2}\mathrm{I}_{n/2-1}(\kappa)}
\exp(\kappa \mathbf{x}_0^T \mathbf{x})
\end{equation*}
where $\kappa\geq 0$, $\Vert \mathbf{x}_0\Vert =1$ and $\mathrm{I}_m(z)$ is the modified Bessel 
function of the first kind. The parameter $\kappa$ may be thought of as analogous to the 
reciprocal of the variance of the Gaussian distribution and $\mathbf{x}_0$ as the mean direction 
about which the points $\mathbf{x}$ cluster.
We choose initial data with average momentum $0$ where 
$\left.j^0\right|_{t=0}=f_3(\mathbf{x};\,\left.\mathbf{x}_0\right|_{\theta=\pi/3,\phi=\pi/3},\, 96)$ 
and $j^1=j^2=0$. 
For the mass parameter we take $\mu=1.2$ and evolve Eq.\eqref{eq:DiracEOM1} and 
Eq.\eqref{eq:DiracEOM2} on the temporal range $t\in[0,\,0.99]$.
As before we use explicit RK4 as the temporal integrator and upon performing SCCT (Fig.~\ref{fig:eomFRWfieldcvgce}) 
find $4$th order convergence in time as expected.
Once again the continuity equation must hold and probability must be conserved. 
These properties are checked for the present numerical calculation in 
Fig.~\ref{fig:contin_frw_sub1} and Fig.~\ref{fig:prob_frw_sub2}.
For each chosen band-limit we observe growth in numerical error of several orders of magnitude as $t\rightarrow0.99$. 
This is reasonable when we take into account that the spatial geometry is shrinking 
to a point ($a(t)\rightarrow0$ as $t\rightarrow 0.99$).
We now fix the band-limit as $L=129/2$ together with temporal grid
$N_\mathrm{st}=6.4\times10^4$.  For the aforementioned resolutions snapshots of
the dynamics are displayed on an Aitoff-Hammer projection~\cite{snyder:1987map}
in Fig.~\ref{fig:DiracFRWSoln}.  In Fig.~\ref{fig:DiracFRWSoln_a} we see that
the probability density $j^0$ is initially Gaussian in character and
subsequently due to the rotational symmetry of the initial state about the mean
direction $\mathbf{x}_0$ we observe evolution towards a ring-like structure
indicating dynamics governed by dispersion (Fig.~\ref{fig:DiracFRWSoln_b}).  In
Fig.~\ref{fig:DiracFRWSoln_c} we see evolution towards the antipode of
$\mathbf{x}_0$, however we find that this does not coincide with a
reconstruction of the initial condition in the limit $t\rightarrow 1$
(Fig.~\ref{fig:DiracFRWSoln_d}).  Observe that the average amplitude of
$\lim_{t\rightarrow 0}j^0(t)$ increases dramatically; this is to be expected due
to conservation of probability $Q$ and the form of the volume element for this
particular geometry used in its calculation (Eq.\eqref{eq:chargeSub}).

\section{Conclusion}\label{sec:conc}
We have presented a new spectral algorithm for half-integer spin-weighted
functions on $\mathbb{S}^2$ based on spin-weighted spherical harmonics (SWSH) by
extending the method for the integer case presented in
\cite{Huffenberger:2010hh,numericalEvolutions2014Beyer}.  Our implementation of
this spectral algorithm shows excellent agreement with the theory we have
presented. Indeed we find that the numerical error scaling comparable to the
integer SWSH algorithm \cite{Huffenberger:2010hh} and that exponential
convergence properties are displayed as anticipated. Furthermore, the expected
algorithmic complexity of $\mathcal{O}(L^3)$ is retained.  In addition, we have
outlined how one may construct the $2+1$ Dirac equation on a curved space-time,
adapting the result to the $\eth$-formalism and for a geometry with spatial
topology of $\mathbb{S}^2$.  Viewing the Dirac equation as an IVP we
demonstrated how a pseudo-spectral approach can be combined with the
half-integer SWSH algorithm and analysed several different situations with
distinct spatial geometries, examples including linear interpolation between
distinct spatial configurations with time and the case of an imploding $2+1$
FRW-like model.  The numerical solutions that we constructed obey anticipated
convergence rates for the temporal solver used (explicit RK4) and physical
invariants (current continuity and probability) are shown to be preserved during
the time-evolution to an excellent degree.

\section{Acknowledgment}
\label{sec:acknowledgment}

JF is grateful to G. Sparling for several enlightening discussions. This research was partly
supported by the Marsden Fund of the  Royal Society of New Zealand.

\appendix

\section{The Dirac equation on a 3-dimensional Lorentz manifold}
\label{sec:dirac-equation-3}

Let $\mathbb{V}$ be a $n$-dimensional real vector space equipped with a quadratic
form $\eta$ with signature $n-2$, i.e., $\eta$ can be represented in diagonal
form as $\eta=\mathrm{diag}(1,-1,\ldots,-1)$. Let $\mathfrak{c}(\mathbb{V},\eta)$
be the Clifford algebra associated to $(\mathbb{V},\eta)$. This is the unique
algebra with unit $\mathbf{1}$ and so called structure map $\gamma: \mathbb{V} \to
\mathfrak{c}(\mathbb{V},\eta)$ such that for every $v\in\mathbb{V}$ the relation
$\gamma(v) \gamma(v) = \eta(v,v) \mathbf{1}$ holds. Let $(e_1,\ldots,e_n)$ be a
basis of $\mathbb{V}$ and define $\eta_{ab}=\eta(e_a,e_b)$ and
$\gamma_a=\gamma(e_a)$. Then, by polarization of the defining relation, one finds
the well-known Clifford-Dirac anti-commutator relations
\[
\{\gamma_a,\gamma_b\} = \gamma_a\gamma_b + \gamma_b\gamma_a = 2\, \eta_{ab} \mathbf{1}.
\]
Irreducible representations $S$ of the Clifford algebras have dimensions $N=2^{n/2}$
for $n$ even and $N=2^{(n-1)/2}$ for $n$ odd. It is well known~(see
\cite{Penrose:1986tf}) that these representations can be constructed recursively
from the lower dimensional ones. The representation space $S$ of the Clifford
algebra $\mathfrak{c}(\mathbb{V},\eta)$ is called the spin space. Depending on
the particular case, the spin space is equipped with certain invariant
structures, see~\cite{Penrose:1986tf},\cite{Porteous95:clifford},\cite{Sparling:XXvo} for
detailed discussions.

The commutators of the generators
$\gamma_{ab}:=\gamma_{[a}\gamma_{b]} = \frac12 \left[\gamma_a,\gamma_b\right]$
generate a Lie algebra with commutation relations
\[
\left[\gamma_{ab},\gamma_{cd}\right] = \eta_{cb}\gamma_{ad} -
\eta_{ca}\gamma_{bd} - \eta_{db}\gamma_{ac} + \eta_{da}\gamma_{bc}.
\]
These are the commutation relations for the Lie algebra $\mathfrak{o}(\eta)$ of
the (pseudo-) orthogonal group $O(\eta)$ associated with the bilinear form $\eta$. This Lie
algebra acts on the subspace $\gamma[\VV]\subset \mathfrak{c}(\mathbb{V},\eta)$
as follows: for any skew bi-vector $w^{ab}$ and every vector $v^a$ we define
$\gamma(w):=w^{ab}\gamma_{ab}$ and $\gamma(v)=v^a\gamma_a$, then the action of $\mathfrak{o}(\eta)$ is
via commutation in  $\mathfrak{c}(\mathbb{V},\eta)$
\[
\mathfrak{o}(\eta)\times \gamma[\VV] \to \gamma[\VV],\quad(\gamma(w), \gamma(v)) \mapsto \gamma(w)\gamma(v) - \gamma(v)\gamma(w).
\]
We can see the corresponding action on $\VV$ from the explicit calculation
\[
\gamma(w)\gamma(v) - \gamma(v)\gamma(w) = w^{ab}v^c \left(\gamma_{ab}\gamma_c -
  \gamma_c \gamma_{ab}\right) = w^{ab}v^c\left(\eta_{bc}\gamma_a -
  \eta_{ac}\gamma_b \right) = 2 w^a{}_bv^b\gamma_a.
\]
Thus, the action on $\gamma[\VV]$ corresponds to the linear mapping defined by
$2w^a{}_b$ on $\VV$, which is clearly anti-symmetric with respect to $\eta$. The same
Lie algebra also acts on the spin-space~$S$ via the representation of the
Clifford algebra
\[
\mathfrak{o}(\eta)\times S \to S,\quad(\gamma(w), \psi) \mapsto \gamma(w)\psi.
\]

In order to make use of spinors on the manifold $\sM$ we associate at every
point $x\in\sM$ the Clifford algebra $\mathfrak{c}(T_x\sM,g_x)$ and its
representation space which we denote by $S_x$. The rigorous construction is
invariantly described in terms of (associated) vector bundles but we will not go
into the details here. Instead we refer to~\cite{Lawson:1990} for a complete
description\footnote{Note, that in the general case there are global
topological obstructions to the existence of these bundles, see~\cite{Lawson:1990,Penrose:1986tf}. However, since we
are interested mainly in the 3-dimensional case where these obstructions do not
exist we will not discuss them here.}. The collection of
all spin spaces $S_x$ forms the spin bundle $S(\sM)$ over $\sM$ and sections of
this bundle are referred to as spinor fields or simply spinors. Similarly, we
obtain the Clifford bundle as the collection of all Clifford algebras
$\mathfrak{c}(T_x\sM,g_x)$. The structure maps defined at every point $x$ yield
a bundle map $\gamma$ from the tangent bundle $T\sM$ to the Clifford bundle,
assigning to every tangent vector $v\in T_x\sM$ an element $\gamma_x(v)$ in the
Clifford algebra at $x$.

The connection $\nabla$ on $\sM$ can be lifted to a connection also denoted by
$\nabla$ on $S(\sM)$ (see~\cite{Trautman:2008gn}). It can be uniquely characterized by the fact that the
structure map $\gamma$ and the invariant structures on the spin spaces are
covariantly constant. With this connection one can define the Dirac operator on
$\sM$ as follows: we pick a basis $(e_1,\ldots,e_n)$ of the tangent space
$T_x\sM$ at each $x\in\sM$. Now the anti-commutator relations are
\[
\{\gamma_a,\gamma_b\} =  2 g_{ab} \mathbf{1}
\] 
which hold at every $x\in\sM$ with the appropriate definition of $g_{ab}$
and $\gamma_a$.  The Dirac operator $\DiracD$ acting on spinor
fields $\psi$ is defined by
\begin{equation}
\DiracD \psi := g^{ab}\gamma(e_a) \nabla_{e_b}\psi = \gamma^a
\nabla_{e_a}\psi.\label{eq:DiracD}
\end{equation}

In this paper we assume the vector space $\VV$ to have dimension 3 and to be
equipped with a metric $\eta$ with signature $(+,-,-)$. The corresponding
Clifford algebra $\mathfrak{c}(\VV,\eta)$ has a
4-dimensional real representation~\cite{Porteous95:clifford,Sparling:XXvo}, which may
conveniently be described within $M(2,\CC)$ as the real algebra generated by the
`Dirac matrices'
\[
\gamma_0 = 
\begin{bmatrix}
  -1 & 0\\
  0 & 1
\end{bmatrix},
\qquad
\gamma_1 = 
\begin{bmatrix}
  0 & 1\\
 -1 & 0
\end{bmatrix},
\qquad
\gamma_2 = 
\begin{bmatrix}
  0 & \ii\\
  \ii & 0
\end{bmatrix}.
\]
The spin space is $\CC^2$ (regarded as a 4-dimensional real vector space) and
carries a sesquilinear form (Hermitean inner product) defined by
\[
\langle\psi,\phi\rangle:= \psi^*\gamma_0\phi,
\]
where $\psi^*$ is the Hermitean conjugate of $\psi$. For given spinors $\psi=\left[
\begin{smallmatrix}
  \psi_1\\\psi_2
\end{smallmatrix}\right]
$ 
and
$\phi=\left[
\begin{smallmatrix}
  \phi_1\\\phi_2
\end{smallmatrix}\right]
$ 
this product is
\[
\langle\psi,\phi\rangle = -\bar{\psi}_1\phi_1 + \bar{\psi}_2\phi_2.
\]
The generators are symmetric with respect to this product, i.e., we have
\[
\langle\psi,\gamma_a \phi\rangle = \langle\gamma_a \psi,\phi\rangle .
\]
Given two spinors $\phi$ and $\psi$ there is a naturally defined covector
$\alpha=\alpha(\phi,\psi)$ with components $\alpha_a =
\langle\phi, \gamma_a\psi\rangle$. When $\phi=\psi$, then this covector is 
\[
\alpha = \left[  |\psi_1|^2 + |\psi_2|^2,  -\bar{\psi}_1\psi_2 - \bar{\psi}_2\psi_1,  \ii (\bar{\psi}_2 \psi_1 - \bar{\psi}_1 \psi_2) \right].
\]
Hence, it is real and null whenever $\langle\psi,\psi\rangle=0$.

The commutators of the Dirac matrices $\gamma_a$ are infinitesimal generators of the spin
group $\mathrm{Spin}(1,2)$. They can be written here in form
\[
\gamma_a\gamma_b - \gamma_b\gamma_a = 2\ii \epsilon_{ab}{}^c \gamma_c.
\]

Choosing an orthonormal frame $(e_0,e_1,e_2)$ on $\sM$ we can express the Dirac operator on
$\sM$ with respect to the chosen frame explicitly. To this end we need the Ricci
rotation coefficients for the frame defined by
\[
\nabla_{e_a}e_b = \Gamma_{ab}{}^c e_c.
\]
The condition that the structure map be covariantly constant translates into the
equation
\[
\Gamma_{ab}{}^c \gamma_c = \Upsilon_b \gamma_a - \gamma_a \Upsilon_b
\]
where the $\Upsilon_b$, which depend on the point $x\in\sM$, are endomorphisms
of the spin space $S_x$. More precisely, they are linear combinations of the
infinitesimal generators of $\mathrm{Spin}(1,2)$. This equation can be
solved uniquely for the $\Upsilon$ and yields 
\[
\Upsilon_a = -\frac{\ii}2 \Gamma_{ba}{}^c\epsilon_c{}^{bd}\gamma_d .
\]
The connection defined in this way also leaves the complex structure and the
sesquilinear product invariant. With these `spin coefficient' matrices we can
express the Dirac operator acting on an arbitrary spinor $\psi$ as
\[
\DiracD \psi = \eta^{ab}\gamma_a  \left(e_b(\psi) + \Upsilon_b \psi\right).
\]

The Dirac equation on $\sM$ can be obtained from a variational principle. For
any given spinor field $\psi$ with compact support on $\sM$ we can write down
the action functional 
\[
\mathfrak{A}[\psi]:=\int_\sM \ii \langle\psi,\DiracD\psi\rangle + \mu \,\langle\psi,\psi\rangle\, d\sM,
\]
where here and later on $d\sM$ denotes the invariant volume form on a manifold
$\sM$. This defines a real number since the imaginary part can be rewritten as
\[
\ii\!\int_\sM  \langle\psi,\DiracD\psi\rangle + \langle\psi,\DiracD\psi\rangle \, d\sM
=
\ii\!\int_\sM  \nabla_a\alpha^a \,  d\sM = 0.
\]
Variation with respect to $\psi$ yields the Dirac equation
\[
\ii \DiracD\psi + \mu \psi = 0.
\]
To every solution $\psi$ of the Dirac equation the covector $\alpha_a$ defines a conserved current
$j^a$ denoted by
\[
j^a = \langle\psi,\gamma^a \psi\rangle.
\]
Taking the divergence of this equation we find
\begin{equation}
  \begin{aligned}
    \nabla_a j^a &= \langle\nabla_a\psi,\gamma^a \psi\rangle +
    \langle\psi,\gamma^a \nabla_a\psi\rangle \\ &= \langle\DiracD\psi,\psi\rangle +
    \langle\psi,\DiracD\psi\rangle = \langle\ii\mu\psi,\psi\rangle +
    \langle\psi,\ii\mu\psi\rangle = 0.
  \end{aligned}
\label{eq:divcurspcpt}
\end{equation}
With this current we can define a conserved quantity in the usual way. Let $t$
be a time coordinate for $\sM$ defined on an interval $I=[0,T]$ for some
arbitrary $T$ and let
$\Sigma$ be a 2-dimensional manifold. We assume that $I\times\Sigma$ is embedded
into $\sM$ as a 3-dimensional submanifold $\sV$ via the embedding
$i:I\times\Sigma \hookrightarrow \sM$. Then $i_t:\Sigma\to\sM, x \mapsto (t,x)$
embeds $\Sigma$ as a space-like hyper-surface $\Sigma_t$ into $\sM$. Let
$\sS =\partial \Sigma$ be the boundary of $\Sigma$, which $i_t$ embeds as a
1-dimensional submanifold $\sS_t$ into $\sM$. Now the boundary of $\sV$ is
$\sV=\Sigma_0 \cup \sT \cup \Sigma_T$ where $\sT=\bigcup_{t\in I} \sS_t$.

Integrating the divergence equation over $\sV$ and using the generalised Stokes'
theorem we obtain
\[
0 = \int_\sV \nabla_aj^a\,d\sV = \int_{\partial\sV} j^a \,d\sV_a ,
\]
where we denote the hyper-surface forms on the boundary $\partial\sV$ by
$d\sV_a$. Introducing the future-pointing time-like normal $t_a$ to the hyper-surfaces
$\Sigma_t$ and the outward normal $n_a$ to $\sT$ we can write
\[
\int_{\Sigma_0} j^at_a \,d\Sigma = \int_{\Sigma_T} j^at_a \,d\Sigma +  \int_{\sT} j^an_a \,d\sT. 
\]
Defining the scalar ``charge'' (also interpreted as probability) $Q$ and its flux $J$ by
\begin{equation}
Q(t) = \int_{\Sigma_t} j^at_a \,d\Sigma ,\qquad
J(t) = \int_{\sS_t} j^an_a \,d\sS\label{eq:chargeflux}
\end{equation}
we can write the balance law
\begin{equation}
Q(T) = Q(0) + \int_0^T J(t)\, dt\label{eq:balance}
\end{equation}
which holds for arbitrary values of $T$. With appropriate boundary conditions
(for instance, when $\Sigma$ has no boundary as we assume below)
for the Dirac field one can make the flux vanish so that the balance law turns
into the conservation equation for $Q$
\[
Q(T) =  Q(0).
\]

We now specialize to the case $\sM\sim \mathbb{R} \times \mathbb{S}^2$ with the
metric 
\begin{equation}
g=\mathrm{d}t\otimes\mathrm{d}t-\mathcal{F}^{-2}(t,\vartheta,\varphi)
\left(\mathrm{d}\vartheta\otimes\mathrm{d}\vartheta
+\sin^2\vartheta\,\mathrm{d}\varphi\otimes\mathrm{d}\varphi \right),
\end{equation}
where $(\vartheta,\varphi)$ are standard polar coordinates for the 2-sphere. The
function $\mathcal{F}(t,\vartheta,\varphi)$ is a conformal factor relating the
induced metric at every instant of time $t$ on $\mathbb{S}^2$ to the standard metric of
the unit 2-sphere. We choose the orthonormal frame as
\begin{equation}
e_0 = \partial_t,\quad 
e_1 = \mathcal{F}\, \partial_\vartheta ,\quad 
e_2 = \mathcal{F}\csc\vartheta\, \partial_\varphi. \label{eq:frames}
\end{equation}
The non-vanishing Ricci rotation coefficients are 
\begin{equation}
\Gamma_{11}{}^0 = - \frac{\dotF}\F, \quad 
\Gamma_{22}{}^0 = - \frac{\dotF}\F, \quad 
\Gamma_{12}{}^1 = - \frac{\F_\varphi}{\sin\vartheta}, \quad 
\Gamma_{22}{}^1 = - \left(\F_\vartheta - \F \cot\vartheta \right).  \label{eq:riccicoeffs}
\end{equation}
From these we obtain the non-vanishing `spin coefficient' matrices
\begin{equation}
\Upsilon_1 = \frac\ii2
\begin{bmatrix}
  {\F_\varphi/\sin\vartheta} & \ii \dotF/\F\\
  \ii {\dotF}/{\F} & -{\F_\varphi}/{\sin\vartheta} 
\end{bmatrix},\quad
\Upsilon_2 = \frac\ii2
\begin{bmatrix}
  \F_\vartheta - \F\cot\vartheta & -{\dotF}/{\F}\\
  {\dotF}/{\F} & -\F_\vartheta + \F\cot\vartheta
\end{bmatrix}.\label{eq:spincoeffs}
\end{equation}
In order to make use of the spin-weighted formalism we need to identify the
spin-weights of the quantities involved. To this end we consider the
infinitesimal rotations of tangent vectors to the sphere as induced by the
$(1,2)$-Lorentz group $O(\eta)$ at every point on $\sM$. Consider the
infinitesimal rotation given by $\frac12\gamma_{12}=\frac\ii2 \gamma_0$ acting
as follows on the two frame vectors on the sphere
\[
\gamma(e_1) = \gamma_1 \mapsto \frac12 \left[\gamma_{12},\gamma_1\right] =
\gamma_2,
\quad
\gamma(e_2) = \gamma_2 \mapsto \frac12 \left[\gamma_{12},\gamma_2 \right] =
-\gamma_1. 
\]
In terms of the complex null-vector $m = \frac1{\sqrt2} (e_1-\ii e_2)$ this is
\[
\gamma(m) \mapsto \ii\, \gamma(m),
\] 
so $\frac12\gamma_{12}$ generates the frame rotations
$m \mapsto e^{\ii\alpha}\,m$. But it also acts on the spin-space at each point:
for every spinor $\psi = [\psi_1, \psi_2]^T$ we have
\[
\frac12\gamma_{12}\psi = \frac\ii2 \gamma_0 
\begin{bmatrix}
  \psi_1\\\psi_2
\end{bmatrix}
=
\frac\ii2\begin{bmatrix*}[r]
  -\psi_1\\ \psi_2
\end{bmatrix*}.
\]
Therefore, assigning spin-weight $+1$ to $m$ results in the spin-weights
$-\frac12$ and $+\frac12$ for $\psi_1$ and $\psi_2$, respectively. Henceforth, we denote
the spinor $\psi$ therefore by $\psi = [\psi_-,\psi_+]^T$. Since $\F$ is a
scalar function it has spin-weight 0.

Now we are in a position to write down the Dirac
equation. Using~\eqref{eq:DiracD}, \eqref{eq:frames}, \eqref{eq:riccicoeffs} and
\eqref{eq:spincoeffs} we find the two equations
\begin{align}
  \partial_t \psi_- &= -\ii \mu \psi_- - \frac12 \eth'\F\, \psi_+ - \F\, \eth'
  \psi_+ + \frac{\F_t}\F\,\psi_- ,\\
  \partial_t \psi_+ &= \ii \mu \psi_+ - \frac12 \eth\F\, \psi_- - \F\, \eth
  \psi_- + \frac{\F_t}\F\,\psi_+.
\end{align}

The probability or charge density is 
\[
j^0 = |\psi_+|^2 + |\psi_-|^2.
\]
We take $\sV$ in the form $(0,t)\times \mathbb{S}^2$ with hyper-surfaces of constant time
$\Sigma_\tau\sim \mathbb{S}^2$ then the `charge' integral as defined in \eqref{eq:chargeflux} yields the
expression
\[
Q(\tau) = \int_{S^2} |\psi_+(\tau,\vartheta,\varphi)|^2 + |\psi_-(\tau,\vartheta,\varphi)|^2\;\frac{\sin\vartheta\,\dd\vartheta\,\dd\varphi}{\F^2(\tau,\vartheta,\varphi)}.
\]
Since $\Sigma_\tau\sim \mathbb{S}^2$ has no boundary, \eqref{eq:balance} implies that for every $\tau \in (0,T)$
\begin{equation}\label{eq:chargeSub}
Q(\tau) = Q(0) = \mathrm{const}.
\end{equation}

\end{document}